\documentclass[linenumbers]{emulateapj}
\usepackage[backref, breaklinks, colorlinks, citecolor=blue, linkcolor=magenta]{hyperref}
\usepackage[dvipsnames]{xcolor}

\shorttitle{The Scatter Matters}
\shortauthors{Sanchez et. al.}

\usepackage{natbib}
\begin{document} 

\title{The Scatter Matters: Circumgalactic Metal Content in the Context of the $M-\sigma$ Relation}

\author{N. Nicole Sanchez\altaffilmark{1,2,3}}
\author{Jessica K. Werk\altaffilmark{3}}
\author{Charlotte Christensen\altaffilmark{4}}
\author{O. Grace Telford\altaffilmark{5}}
\author{Michael Tremmel\altaffilmark{6,7}}
\author{Thomas Quinn\altaffilmark{3}}
\author{Jennifer Mead\altaffilmark{8}}
\author{Ray Sharma\altaffilmark{5}}
\author{Alyson Brooks\altaffilmark{5,9}}

\affil{$^1$The Observatories of the Carnegie Institution for Science, 813 Santa Barbara Street, Pasadena, CA 91101, USA, nsanchez@carnegiescience.edu}
\affil{$^2$Cahill Center for Astronomy and Astrophysics, California Institute of Technology, MC249-17, Pasadena, CA 91125, USA}
\affil{$^3$Astronomy Department, University of Washington, Seattle, WA 98195, USA}
\affil{$^4$Physics Department, Grinnell College, 1116 Eighth Ave., Grinnell, IA 50112, USA}
\affil{$^5$Rutgers University, Department of Physics and Astronomy, 136 Frelinghuysen Road, Piscataway, NJ 08854, USA}
\affil{$^6$Astronomy Department, Yale University, P.O. Box 208120, New Haven, CT 06520, USA}
\affil{$^7$School of Physics, University College Cork, College Road, Cork T12 K8AF, Ireland}
\affil{$^8$Columbia University, Department of Astronomy, New York, NY 10025, USA}
\affil{$^9$Center for Computational Astrophysics, Flatiron Institute, 162 5th Ave, New York, NY 10010, USA}

\begin{abstract}\label{abs:abstractlabel}
 
The interaction between supermassive black hole (SMBH) feedback and the circumgalactic medium (CGM) continues to be an open question in galaxy evolution. In our study, we use SPH simulations to explore the impact of SMBH feedback on galactic metal retention and the motion of metals and gas into and through the CGM of L$_{*}$ galaxies. We examine 140 galaxies from the 25 Mpc cosmological volume, {\sc Romulus25}, with stellar masses between 3 $\times$ 10$^{9}$ -- 3 $\times$ 10$^{11}$ M$_{\odot}$. We measure the fraction of metals remaining in the ISM and CGM of each galaxy, and calculate the expected mass of its SMBH based on the $M-\sigma$ relation. The deviation of each SMBH from its expected mass, $\Delta M_{BH}$ is compared to the potential of its host via $\sigma$. We find that SMBHs with accreted mass above the empirical $M-\sigma$ relation are about 15\% more effective at removing metals from the ISM than under-massive SMBHs in star forming galaxies. Over-massive SMBHs suppress the overall star formation of their host galaxies and more effectively move metals from the ISM into the CGM. However, we see little evidence for the evacuation of gas from their halos, in contrast with other simulations. Finally, we predict that C IV column densities in the CGM of L$_{*}$ galaxies may depend on host galaxy SMBH mass. Our results show that the scatter in the low mass end of $M-\sigma$ relation may indicate how effective a SMBH is at the local redistribution of mass in its host galaxy.

\end{abstract}


\section{Introduction} 
\label{sec-intro}


The vastly different scales between the event horizon of a supermassive black hole (SMBH) and the size of its host galaxy have been evocatively described by \cite{Savorgnan2016} as the difference between a grain of sand and the entirety of the Saharan Desert. (A difference of approximately 10 orders of magnitude). While the size difference between these objects makes their interaction surprising, mounting evidence continues to connect the evolution and properties of galaxy hosts to their SMBHs \citep{Magorrian1998,Haehnelt1998a,Ferrarese2000,Gebhardt2000,Reines2015a,Saglia2016a}. 

The relation between the mass of the central supermassive black hole, M$_{BH}$, and the stellar dispersion of its host galaxy's bulge, $\sigma_{*}$, is one of the most fundamental relations drawn between the SMBH and its host galaxy \citep[][and citations therein]{Kormendy2013}. Colloquially known as the $M-\sigma$ relation, this observed relation shows a tight correlation across three orders of magnitude in SMBH mass and is theorized to tie together the growth of a SMBH--during its tenure as an active galactic nucleus (AGN)--and the winds launched from its accretion disk. These winds are responsible for removing some of the gas necessary for continued star formation in the galaxy. In this way, the energetics of the SMBH work to regulate the star formation in the bulge of massive galaxies. When the SMBH is no longer accreting or driving outflows, gas accretion and star formation can resume \citep{Alexander2005a,Papovich2006,Volonteri2012}. The quantity $\sigma_{*}$ not only reflects the mass of its host galaxy, but also approximates the depth of the galaxy's potential well \citep{Ferrarese2000,Zahid2018,Ricarte2019}. 

The scatter in the $M-\sigma$ relation can further illuminate the processes that drive galaxy evolution at all galaxy masses. At the high mass end, there is less scatter and it is dominated by more massive BHs residing in massive ellipticals above $\sim$ 10$^{13}$ M$_{\odot}$ \citep{VanDenBosch2007,Moster2010,Natarajan2011,Emsellem2011,Dubois2015}. However, lower mass BHs live in a more diverse range of galaxy masses resulting in scatter that is more pronounced on the low mass end of the relation. This low-mass-end scatter may be explained by variable pathways that drive SMBH growth \citep{Micic2007a,Volonteri2009a,Reines2013a,Graham2015,Sharma2019}. Galaxy mergers are thought to fuel SMBHs, in addition to building up galaxies and contributing to the assembly of bulges \citep{DiMatteo2005,Shen2008,Sanchez2017}. 
Furthermore, episodes of SMBH fueling result in feedback that removes gas from the galaxy suppressing both SMBH growth and future star formation \citep{Schawinski2010,Pontzen2017,Sanchez2021}. A concerted effort has been put forth to explain the physical processes that result in the scatter on the $M-\sigma$ relation. However, the impact on galaxy properties by SMBHs which deviate from the $M-\sigma$ relation has not been well constrained, especially within the context of the circumgalactic medium (CGM).


The rise of observational surveys of the CGM with the Cosmic Origins Spectrograph (COS) on HST has inspired a range of theoretical studies focused on the connection between feedback processes and the content of the CGM. Simulations were initially hard-pressed to match the observational survey results carried out with the Cosmic Origins Spectrograph \citep[e.g.][]{Tumlinson2013,Werk2013}. Predictions for column densities of high ions like \ion{O}{6} were too low, and low ion column densities were difficult to replicate in the simulated environments of cosmological volumes \citep{Oppenheimer2016,Suresh2017}.

More recently, cosmological simulations have updated the subgrid prescriptions in their codes to better characterize the low density, multiphase medium of the CGM \citep{Stinson2012,Shen2012,Vogelsberger2014a,Schaye2015,Tremmel2017}; furthermore, recent work has focused on connecting the impact of energetic feedback from a galaxy's SMBH to the diffuse CGM. Broadly, simulations have shown that the SMBH can impact the CGM in a multitude of ways: heating and evacuating (or removing) gas in the disk to quench star formation in the galaxy \citep[IllustrisTNG and EAGLE simulations,][]{Suresh2017,Nelson2018b,Oppenheimer2016}, driving metal rich gas out of galaxy centers and moving metal-rich gas into (enriching) the CGM \citep[IllustrisTNG and ROMULUS25,][]{Nelson2019,Sanchez2019}, as well as ejecting CGM gas out into the IGM \citep[EAGLE,][]{Oppenheimer2018}. Furthermore, \cite{Mitchell2020} finds that more gas flows out of the halo virial radius than from the ISM of central galaxies in the EAGLE simulations implying increased mass loading within the CGM, while CGM mass fractions decline after explosive episodes of AGN-driven feedback in galaxies from both EAGLE and IllustrisTNG \citep{Oppenheimer2020a}. At lower masses, \cite{Sharma2022} shows that though SMBHs can drive outflows in some dwarf galaxies, gas doesn't often leave the CGM.

Overall, these cosmological simulations seem to paint a similar picture. Both EAGLE and Illustris/IllustrisTNG simulations predict the evacuation of the CGM in the massive galaxies where SMBH processes dominate. However, this may not be the whole story. \cite{Chadayammuri2022} compares CGM radial profiles from eROSITA and mock X-ray observations from the IllustrisTNG and EAGLE cosmological simulations. They find that the luminosity of the CGM of their observed galaxies is higher than that predicted by the simulations indicating that more explosive AGN feedback prescriptions may over-evacuate the CGM of their galaxies.

Furthermore, recent work from \cite{Davies2020} ties the expulsion of gas by SMBH-driven outflows to the scatter in the halo gas fraction at fixed $M_{200}$ in both the IllustrisTNG and EAGLE simulations. Galaxies with more massive BHs (within a fixed halo mass bin) reside within more gas-poor halos, while galaxies with under-massive BHs retain a higher gas fraction in the CGM as well as show elevated star formation rates. \cite{Davies2020} find that the evacuation of CGM gas by SMBH feedback is a critical step in the morphological evolution and quenching of their galaxies. 
These results point to an intrinsic connection between black hole masses and the evolution of the CGM. We follow this line of investigation to further our understanding of how the deviation of a SMBH's mass from empirical expectations impacts its host halo gas.


In this paper, we examine how deviation in SMBH mass from the empirical $M-\sigma$ relation changes the overall effectiveness of SMBH feedback at moving gas and driving metal flows into and out of the CGM. We explore this change across two orders of magnitude in stellar mass, 3 $\times$ 10$^9$ M$_{\odot}$ \textless $M_{*}$ \textless 3 $\times$ 10$^{11}$ M$_{\odot}$. Our study also includes comparisons between our simulations and observational constraints such as metal retention fractions and makes predictions for ion column density measurements in the CGM of galaxies with known SMBH mass measurements.

This paper is organized as follows: Section 2 introduces our simulations and the galaxy selection process, and in Section 3, we describe and analyze our results. In Section 4, we compare our findings to observed measurements from the literature as well as a set of mock observational data and discuss the broader context for our results and their implications. Finally, in Section 5, we summarize and conclude.


\section{Simulated Galaxy Sample}
\label{sec-model}

\begin{figure}[]
\vspace{-2mm}s
\centerline{\resizebox{1.1\hsize}{!}{\includegraphics[angle=0]{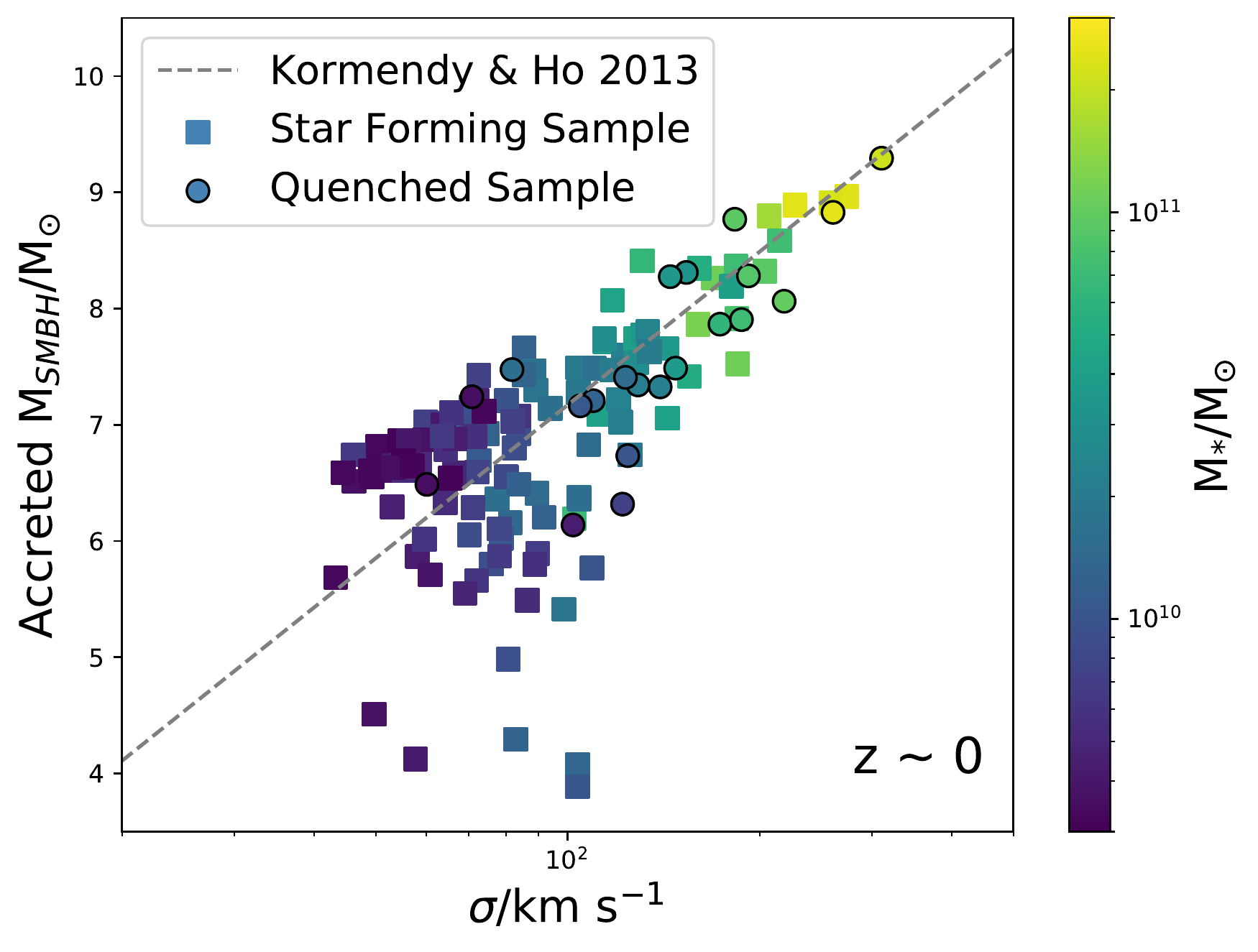}}}
\caption[]{The $M-\sigma$ relation for the 140 galaxies within {\sc Romulus25} that are within our selected stellar mass range and that contain a SMBH. For our measurement of $M_{SMBH}$ we measured the accreted mass for each SMBH to neglect each BH's starting seed mass of $10^6 M_{\odot}$. Star forming galaxies are denoted by squares and quenched galaxies (sSFR \textless 1.6 $\times$ 10$^{-11}$ M$_{\odot}$ yr$^{-1}$) are shown as circles. Points are colored by the stellar mass of the galaxy. The spread of the galaxies fall along the $M-\sigma$ relation, grey dashed line, of \cite{Kormendy2013}, though we note that at the lower mass end our sample tend to lie slightly above the line.}
\label{figure:Msigma}
\end{figure}

\subsection{Simulation Parameters}
\label{subsec:sim_params}

\begin{figure*}[ht!]
\centerline{
\resizebox{0.85\hsize}{!}{\includegraphics[angle=0]{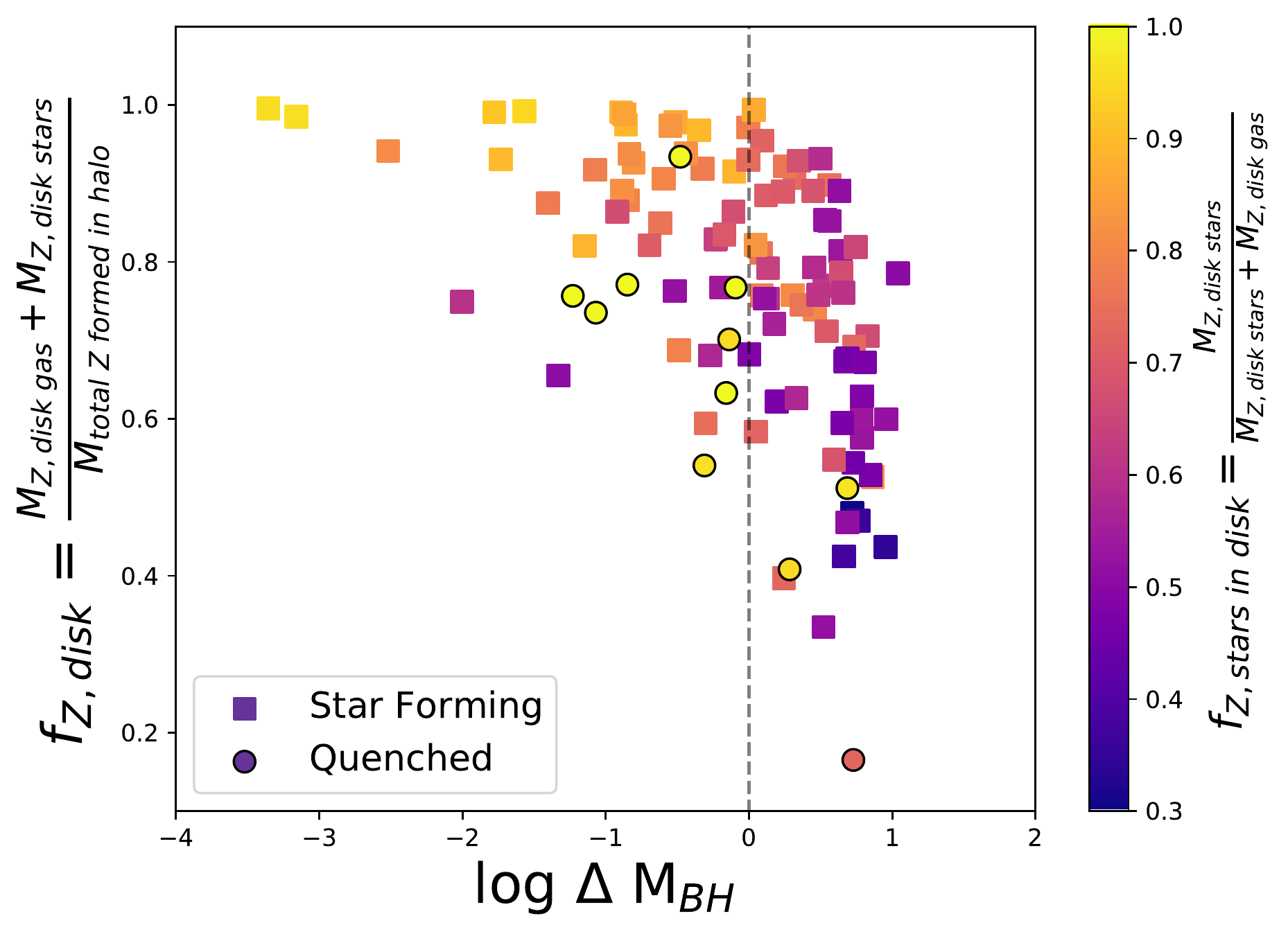}}}
\caption[]{The metal retention, f$_z$, of our sample of galaxies from the {\sc Romulus25} as a function of $\Delta M_{BH}$. Points are colored by the fraction of total disk metals contained in stars. SMBHs that are left of the grey line are under-massive compared to their host galaxy's stellar population (BH masses below the $M-\sigma$ relation, Figure \ref{figure:Msigma}) and maintain more of their metals in their disks or central regions. Meanwhile, galaxies to the right of the line are more effective at removing metals from the disk, with a similar but stronger effect seen in quiescent galaxies (circles).} 
\label{figure:fzdev}
\end{figure*}

All of the galaxies examined in this paper were selected from the {\sc Romulus25} simulation \citep[{\sc R25},][]{Tremmel2017,Ricarte2019,Sharma2019}, a 25 Mpc cosmological volume, run with the smoothed particle hydrodynamics (SPH) N-body tree code, Charm N-body Gravity solver \citep[ChaNGa][]{Menon2015}. ChaNGa adopts its models for cosmic UV background, star formation based on a Kroupa IMF, and `blastwave' supernova feedback from the well-tested GASOLINE code \citep{Wadsley2004,Wadsley2008,Stinson2006,Shen2010}. Rates from supernova Type Ia and Type II are implemented through the Raiteri et. al. (1996) method, using the stellar lifetime calculations of the Padova group for stars with varying metallicities \citep{1993A&AS...97..851A,1993A&AS..100..647B,1994A&AS..106..275B}. We use the following parameters for our stellar subgrid models: star forming efficiency, c$_*$ = 0.15; the fraction of SNe energy that couples to the ISM, $\epsilon_{SN}$ = 0.75; and the amount of SN energy imparted to the gas is 10$^{51}$ ergs. For additional details about the SN ``blastwave'' radius and SN Ia and II metal enrichment prescriptions, see \cite{Stinson2006}.

\begin{figure}[t!]
\centerline{\resizebox{1\hsize}{!}{\includegraphics[angle=0]{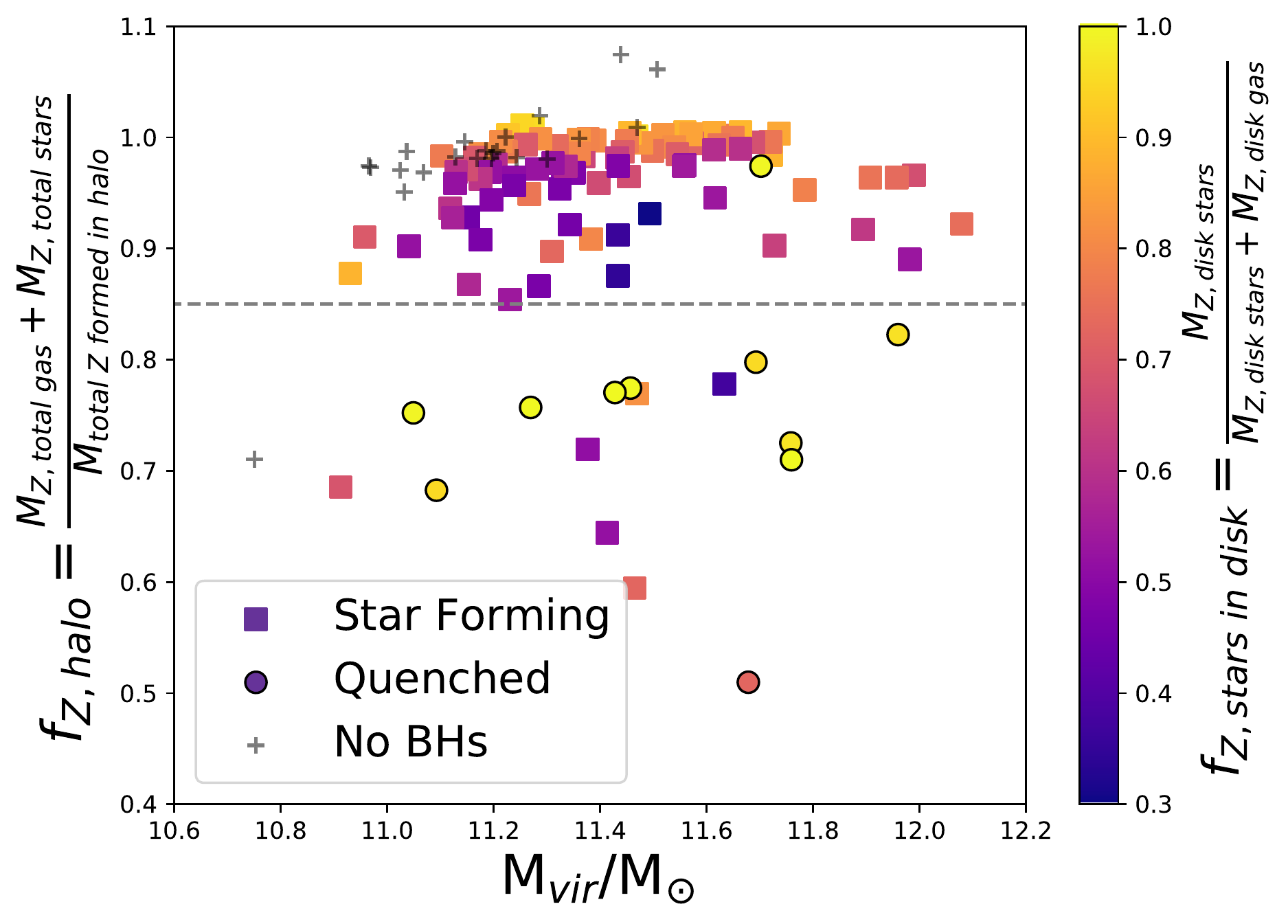}}}
\caption[]{The fraction of all metals retained in the halo as a function of the galaxy halo mass colored by the fraction of disk metals locked in stars. Most of the star forming galaxies (squares) retain at least 85\% of their total formed metals with a small few (6/119 star forming galaxies) losing more. Meanwhile, quenched galaxies (circles) lose anywhere from 0-50\% of their metals. An interesting trend appears in the set of quenched galaxies: more massive galaxies lose fewer metals to the IGM and they appear to have more metals in the gas phase in their central regions (orange circles) compared to their lower mass counterparts. In these lower mass quenched galaxies which lose more metals to the IGM, the metals that remain are almost entirely locked in stars in the disk (yellow circles). Finally, we include the galaxies with no black holes (grey crosses) within R25 that reside within our mass range, and note they are on the lower mass end. They maintain nearly all metals originally formed by the stars in their host galaxies, and in two cases have gained additional metals. 
}
\label{figure:fzcgm_fzdisk}
\end{figure}

ChaNGa includes a SPH formalism which updates the force expression to include a geometric averaged density approach \citep{Wadsley2017}. This hydrodynamics treatment includes thermal and metal diffusion \citep{Shen2010} and reduces artificial surface tension to result in improved resolution of fluid instabilities \citep{Ritchie2001,Menon2015}.

Gas cooling in {\sc R25} is regulated by metal abundance as in \cite{Guedes2011}; however, it does not include a full treatment of metal cooling. 
We include a low temperature extension to the cooling curve which allows gas below 10$^4$ K to cool proportionally to the metals in the gas. Gas above 10$^4$ K cools only by H/He, Bremsstrahlung, and inverse Compton effects \citep[see][for full details]{Tremmel2017}. 

\cite{Shen2012} compare simulations of MW-mass halos at high redshift, finding that with realistic treatments of metal diffusion and stellar IMF that the inclusion of metal-line cooling does not influence the total stellar mass of the galaxy. In other cases, particularly at lower masses, the inclusion of metal-line cooling with the simplistic treatment of star formation at low resolution results in over-cooling, requiring artificially strong feedback to overcome \citep{Christensen2014}. However, it cannot be overlooked that the inclusion of high temperature metal-line cooling can influence the rate at which gas cools out of the CGM onto galaxies \citep{VandeVoort2011b}. At the metallicities ($\sim$10-30\% Z$_{\odot}$) that we expect for the CGM of the most massive halos we study in this work, the lack of metal-line cooling will likely impact the cooling rates by a typical factor of 3\textemdash5 at the peak of the cooling curve at 10$^{5.5-6}$ K when the effect of the UV background is accounted for \citep{Shen2010}. The effect is non-trivial and this does represent a significant caveat to this work that we discuss further in Section \ref{sec:metalcooling}.


{\sc R25} includes updated black hole formation, accretion, and feedback prescriptions. BH formation ties seeds to dense and extremely low metallicity gas to more effectively estimate SMBH populations in a variety of galaxy mass regimes. The SMBH accretion model is based on Bondi-Hoyle, but includes a consideration for angular momentum support from nearby gas. This update allows for more physically motivated growth than Bondi-Hoyle alone \citep{Rosas2016,Angles-Alcazar2017}. Thermal SMBH feedback imparts energy on the nearest 32 gas particles according to a kernel smoothing and is based on accreted mass, $\dot{M}$, via:
\begin{equation}
E = \epsilon_r \epsilon_f \dot{M} c^2 dt,
\end{equation}
where $e_f$ = 0.02 and $e_r$ = 0.1 are the feedback and radiative efficiency, respectively. Accretion is assumed to be constant for one black hole timestep, $dt$. This SMBH feedback prescription has been shown to successfully produce large scale outflows \citep{Pontzen2017,Tremmel2018a}. Finally, an updated dynamical friction prescription has been included to better track SMBH growth and dynamical evolution \citep{Tremmel2015}. For additional details about BH prescriptions, see \cite{Tremmel2017}.

{\sc R25} was run with a $\Lambda$CDM cosmology with $\Omega_0$ = 0.3086, $\Lambda$ = 0.6914, h = 0.67, $\sigma_8$ = 0.77 \citep{Planck2015}. R25 has a Plummer equivalent force softening length of 250 pc and has a UV background through the evolution to $z = 0$ \citep{Haardt2012}. R25 uses a gas and DM particle resolution of 3.4 $\times$ $10^5$ $M_{\odot}$ and 2.1 $\times$ $10^5$ $M_{\odot}$, respectively. Additionally, {\sc R25} has been optimized to match the observed stellar mass\textemdash halo mass relation of \cite{Moster2013} and the SMBH-stellar mass relation using stellar mass and halo mass corrections from \citep{Munshi2013}.

\begin{figure*}[ht!]
\vspace{0mm}
\centerline{\resizebox{1\hsize}{!}{\includegraphics[angle=0]{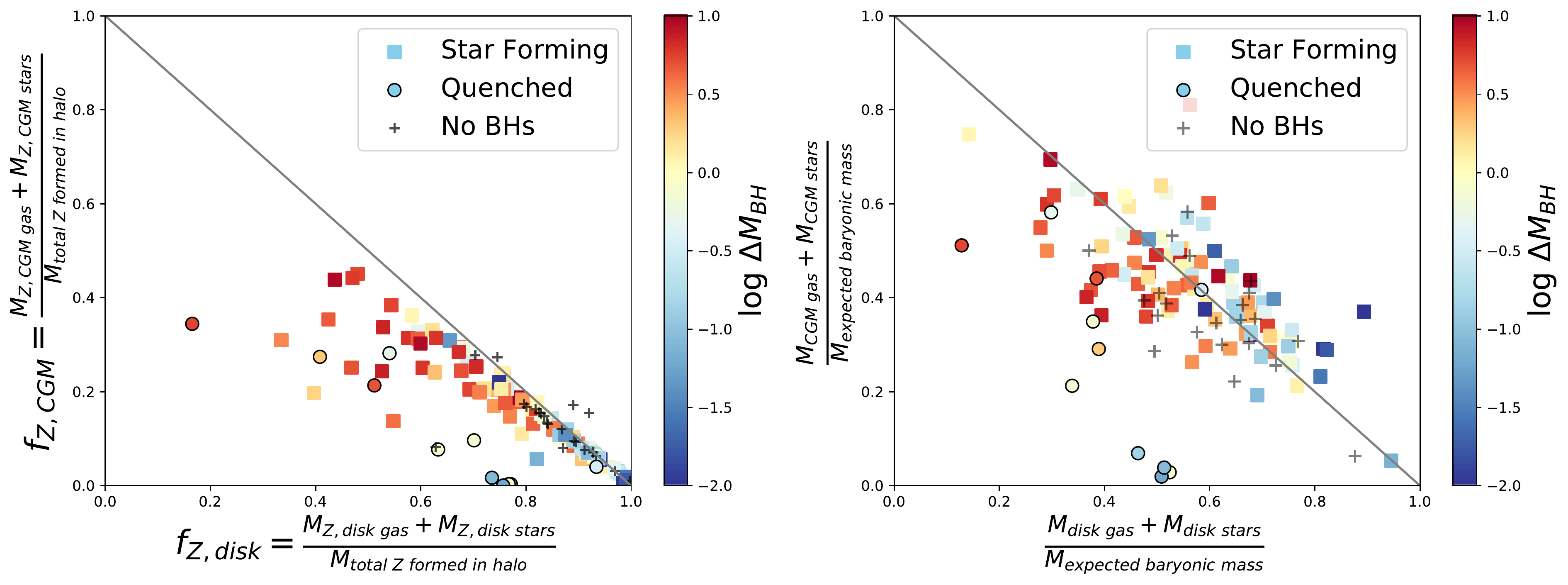}}}
\caption[]{(\textit{Left}) The metal retention of gas and stars in the \textit{CGM} as a function of the metal retention of gas and stars in the \textit{disk} colored by the quantity $\Delta M_{BH}$. The grey solid line indicates the one-to-one line where halos that fall along the line still retain all the metals formed in their galaxies, while galaxies below the line have lost metals to the IGM. This plot shows us that the SF galaxies with over-massive BHs (red squares) lose more metals into the CGM, and in some cases out to the IGM. Meanwhile, galaxies that retain the most metals in their disks host under-massive black holes (blue squares). Galaxies with no black holes appear to occupy a similar range as galaxies with under-massive BHs. All but one of the quenched galaxies (circles) have lost over 20\% of their metals from their central regions. (\textit{Right}) Fraction of total baryonic mass in the CGM as a function of the fraction of total baryonic mass in the disk. Total baryonic fraction is relative to cosmic baryon fraction. Points are colored by $\Delta M_{BH}$. The grey line indicates where the baryonic fraction of the disk and the baryonic fraction of the CGM equals one. We note that nearly all the quenched galaxies fall below this grey line indicating that the loss of disk gas results in the quenching of nearly all of these galaxies. For SF galaxies, we find that galaxies with over-massive black holes (red squares) appear to have more baryons in the CGM compared to the galaxies with under-massive BHs (blue squares). Galaxies with no black holes (grey crosses) lie along the one-to-one line and have a scatter similar to the galaxies with under-massive black holes. Nearly all quenched galaxies (circles) have fewer baryons in the disk than expected.}
\label{figure:fzcgm_fzdisk_fbaryons}
\vspace{3mm}
\end{figure*}

\subsection{Isolated Galaxies with M$_*$ = $10^{9.5-11.5}$ M$_{\odot}$}
\label{subsec:gxy_params}

For comparison with observations, we select our sample of galaxies using a roughly L$^*$ stellar mass range that has been well inspected by observations \citep{Tripp2011,Tumlinson2013,Werk2013,Borthakur2015,Wilde2021}: 3 $\times$ 10$^9$ M$_{\odot}$ \textless\  M$_{*}$ \textless\ 3 $\times$ 10$^{11}$ M$_{\odot}$. Within {\sc R25}, there are 282 galaxies within this stellar mass range at z $\sim$ 0. We further refine our selection to remove galaxies that we considered satellites. We define satellites as galaxies within 300 kpc of another more massive galaxy. Using this definition, our main sample consists of 140 galaxies that host SMBHs at their centers. There are 119 star forming (SF, sSFR \textgreater \ 1.6 $\times$ 10$^{-11}$) galaxies in our sample and 21 are quenched (Q, sSFR \textless  1.6 $\times$ 10$^{-11}$) at $z \sim 0$. In addition, we find 20 isolated galaxies within this mass range that do not host a SMBH at their center.

Figure \ref{figure:Msigma} shows the $M-\sigma$ relation for our full sample of 140 galaxies with central SMBHs. Our sample falls along the line produced using the $M-\sigma$ equation of \cite{Kormendy2013}:
\begin{equation}
\frac{M_{BH}}{10^9 M_{\odot}} = \Big(0.309 \Big) \left(\frac{\sigma}{200\ km\ s^{-1}}\right)^{4.38} 
\label{eq:KandH}
\end{equation}
where $M_{BH}$ represents the mass of the SMBH and $\sigma$ is isolated stellar velocity dispersion of a central bulge, if one exists. To account for the seed mass of 10$^6$ M$_{\odot}$, we will also define $M_{BH,acc}$ as the contribution of SMBH mass obtained by accretion. 

Meanwhile, we calculate the velocity dispersion of each galaxy, $\sigma_{*}$, as in \cite{Ricarte2019}. We select the stellar bulge using a single Sersic profile for each galaxy with an assumed surface brightness cutoff of 32 mags arcsec$^{-2}$ and a maximum radius of 5R${_{half-light}}$. Then we calculate $\sigma_{*}$ from the stars in this selected region using:
\begin{equation}
\sigma = \sqrt{\langle v^2\rangle - \langle v\rangle^2}
\label{eq:sigma}
\end{equation}
where $v$ is the velocity of each particle \citep[see][for further details]{Ricarte2019}. 

As in \cite{Ricarte2019}, the fact that our galaxies mostly lay along the empirical $M-\sigma$ relation indicates that their growth is consistent with the observed phenomenon that SMBHs grow alongside the stellar content of their host galaxies.


\begin{table*}[ht!] 
\caption{Median Metal Retention Values from the {\sc Romulus25} Samples at z $\sim$ 0} 
\centering 
\begin{tabular}{c c c c c} 
\hline\hline 
Sample & \# of Galaxies & $f_{Z,halo}$ $\pm 1\sigma$ & $f_{Z,disk}$ $\pm 1\sigma$  & $f_{Z,CGM}$ $\pm 1\sigma$  \\ [1ex]
\hline 
Star Forming, Over-massive SMBH  & 76 & 0.95 ($\pm$ 0.08) & 0.73 ($\pm$ 0.15) & 0.21 ($\pm$ 0.11)\\ [1ex] 
Star Forming, Under-massive SMBH & 42 & 0.99 ($\pm$ 0.02) & 0.89 ($\pm$ 0.16)& 0.09 ($\pm$ 0.15) \\ [1ex]
Quenched Over-massive SMBH       & 9  & 0.85 ($\pm$ 0.14) & 0.51 ($\pm$ 0.15) & 0.34 ($\pm$ 0.13) \\ [1ex] 
Quenched Under-massive SMBH      & 12 & 0.86 ($\pm$ 0.10) & 0.72 ($\pm$ 0.09) & 0.13 ($\pm$ 0.13) \\ [1ex]
Quenched (All)                  & 21 & 0.85 ($\pm$ 0.12) & 0.63 ($\pm$ 0.16) & 0.26 ($\pm$ 0.15)  \\ [1ex]
$M_*$ \textgreater\  $10^{10}$ M$_{\odot}$       & 76 & 0.97 ($\pm$ 0.07) & 0.76 ($\pm$ 0.17) & 0.16 ($\pm$ 0.15) \\ [1ex] 
$M_*$ \textless\  $10^{10}$  M$_{\odot}$         & 64 & 0.96 ($\pm$ 0.11) & 0.75 ($\pm$ 0.17) & 0.20 ($\pm$ 0.11) \\ [1ex] 
$M_{SMBH}$ \textgreater\  $10^{7.6}$ M$_{\odot}$ & 41 & 0.95 ($\pm$ 0.05) & 0.66 ($\pm$ 0.13) & 0.29 ($\pm$ 0.12) \\ [1ex] 
$M_{SMBH}$ \textless\  $10^{7.6}$ M$_{\odot}$    & 99 & 0.97 ($\pm$ 0.10) & 0.77 ($\pm$ 0.18) & 0.15 ($\pm$ 0.13) \\ [1ex] 
\hline 
\vspace{1mm}
\end{tabular}
\label{table:medians} 
\end{table*}

\section{Results}
\label{sec-results}

The expulsion of metal-rich gas from the center of the galaxy by AGN feedback has recently been shown to be a key process for enriching the CGM \citep{Nelson2019,Sanchez2019}. We continue this line of research to determine whether properties of the galaxy or the SMBH may impact this effect. To do this, we explore the galaxies from {\sc R25} and determine where the metals in each galaxy remain. We split each galaxy into two main components. First, we define the ``disk'' or ``central region'' as the inner 0.1R$_{vir}$ of the galaxy, a conventional definition of the galaxy-CGM boundary \citep{Sales2009,Howk2017}. Then, the CGM is defined to include all particles from 0.1R$_{vir}$ extending out to the virial radius. 

We define the metal retention fraction as the fraction of total metals retained by each component within individual galaxies. For example, the metal retention fraction of the disk/central region component is calculated using the formula from \cite{Telford2019}:
\begin{equation}
f_{Z} = \frac{M_{Z,disk\ gas,present} + M_{Z,disk\ *,present}}{M_{Z,formed}}
\label{eq:fz}
\end{equation}

where M$_{Z,disk\ gas,present}$ and M$_{Z,disk\ *,present}$ are the amount of mass contained in metals in gas and stars within the central region at $z=0$. M$_{Z,formed}$ indicates the amount of metals formed throughout the simulation by stars that reside within the halo at $z = 0$. To calculate this value, we duplicate the calculations done by the simulation using {\sc pynbody} \citep{pynbody} and determine the metal yields from SNIa and SNII for all of the star particles in the halo at $z = 0$.

We calculate M$_{BH,expected}$, the expected SMBH mass for a galaxy's velocity dispersion, using the equation 2 above \cite{Kormendy2013}. From this value, we define the deviation from M$_{BH,expected}$ as:
\begin{equation}
\Delta M_{BH} = M_{BH,accreted} - M_{BH,expected}
\end{equation}
where M$_{BH,accreted}$ is measured from the simulation and is shown on the y-axis of Figure \ref{figure:Msigma}.

From the $\Delta M_{BH}$, we further classify our sample into two sets: galaxies with over-massive SMBHs and galaxies with under-massive SMBHs. Galaxies in the first set, those with over-massive SMBHs compared to their M$_{BH,expected}$, have a positive deviation, $\Delta M_{BH}$ \textgreater 0, and fall above the grey line in Figure \ref{figure:Msigma}. Galaxies with under-massive SMBHs fall below the empirical $M-\sigma$ line and have a negative $\Delta M_{BH}$. We note that in our sample low stellar mass galaxies are more likely to host over-massive black holes. Finally, where applicable, we include the galaxies without central SMBHs for additional comparison.

\subsection{Scatter in the Low Mass End of $M-\sigma$}

In Figure \ref{figure:Msigma}, we see that scatter, and therefore $\Delta$M$_{BH}$, is greatest at the low mass end of the relation as expected by observations \citep{Kormendy2013}. To address the enhanced scatter at the low mass end and remove the impact of galaxy mass from our results, we focus this section (3.1) of our analysis to galaxies in the lower half of our sample with stellar masses below $3 \times 10^{10} M_{\odot}$. Our low mass sample includes 106 from our full sample of 140 galaxies with 95 SF and 11 Q galaxies. All of our galaxies without central SMBHs also fall inside this mass range.

\subsubsection{Metal Retention Correlates with $\Delta$M$_{BH}$}

In Figure \ref{figure:fzdev}, we find that the metal retention in the disk correlates with the deviation of each galaxy's central SMBH from their $M_{BH,expected}$, $\Delta M_{BH}$. Galaxies with over-massive SMBHs can retain significantly less metals within their disks, with star forming galaxies showing a median of f$_{Z,disk}$ = 0.73 of their metals from the disk, and quenched galaxies a median of 0.51 (Table \ref{table:medians}). Galaxies with under-massive black holes retain most, if not all of their metals within the disk, with medians of f$_{Z,disk}$ of 0.89 and 0.72 for star forming and quenched galaxies, respectively. We also find that most of the metals that remain in the disks or central regions of the galaxies with under-massive black holes are locked in stars (yellow points). Meanwhile, galaxies with over-massive SMBHs have a majority of their metal stored in the gas phase (purple points). 

We can determine where the metals lost by each galaxy end up from Figure \ref{figure:fzcgm_fzdisk}, which compares the total metals retained in the entire halo (f$_{z,disk}$ + f$_{z,cgm}$) as a function of halo mass. Each point is colored by the metal mass fraction of the disk/central region in stars as calculated by:
\begin{equation}
f_{z, stars\ in\ disk} = \frac{M_{Z,disk\ stars}}{M_{Z,disk\ stars} + M_{Z,disk\ gas}}
\end{equation}
where each value is calculated as described in Section \ref{sec-model}.

We find that the majority of our star forming galaxies keep nearly all of their metals within the halo with only up to 15\%  of their metals being lost to the IGM. Only $\sim$ 5\% (6/119) of these star forming (SF) galaxies lose more than 15\% of their metals to the IGM (indicated by the dashed grey line at y = 0.85). Furthermore, by comparing the color of the points, we can see that in the star forming galaxies that keep most of their metals, those metals are primarily stored in stars within the disk (yellow-to-orange squares). The SF galaxies that lose more of their metals to the CGM (purple squares) have more of the metals in their disk contained within the gas phase. Thus, this figure could indicate an additional role played by the SMBH. In the SF galaxies with over-massive black holes, the SMBH may not only be responsible for ejecting metals out from the disk, but seems to also play a role in stellar regulation by suppressing overall the star formation and therefore resulting in lower metallicity stars at the host galaxy's center at $z=0$. This result is consistent with \cite{Sharma2022} which looked at the impact of SMBHs on galaxies in the dwarf mass regime, M$_{*}$ = 10$^8$ \textemdash 10$^{10}$. They find that in the galaxies at the high mass end of their sample (M$_{*}$ \textgreater 10$^{9.3}$) SMBHs are driving gas out from the central 0.1R$_{vir}$ into the CGM on fast timescales ($\sim$ 1 Gyr) around z $\sim$ 0.5 \textemdash 1. 

However, Figure \ref{figure:fzcgm_fzdisk} shows distinct differences when we consider the quenched galaxies in our sample. All but one quenched galaxy have lost some metals to the IGM, and nearly half of these have lost at least 15\% of their total metals. Additionally, every quenched galaxy has over 70\% of their metals locked in stars (yellow-to-orange circles). This characteristic likely comes from the fact that these quenched galaxies have lost most if not all of their cold gas by z $\sim$ 0. Thus, the primary contribution of metals come from the stars that remain.


The left panel of Figure \ref{figure:fzcgm_fzdisk_fbaryons} more clearly shows the distribution of metals within each galaxy component. It compares the metal retention in the CGM to the metal retention in the disk. From this figure, we can see the distribution of metals in the components of each galaxy. Galaxies that fall along the grey one-to-one line have maintained all of their metals within the virial radius (a y-value of 1.0 in the left figure). Since we know that most of our galaxies do not lose many metals to the IGM, it is unsurprising that most galaxies fall nearly along this line. Points are colored by each galaxy's $\Delta M_{BH}$ with red points indicating galaxies with the most over-massive black holes and blue points indicating the most under-massive black holes. As expected, the galaxies that retain the most metals in their disks (bottom-right) have under-massive black holes, and from Figure \ref{figure:fzdev}, we know that the metals in their disks are locked primarily in stars. In contrast, the galaxies with over-massive BHs (red) have their disk/central region metals primarily stored in the gas phase in addition to losing more of those metals to the CGM and some to the IGM as well. Thus, the populations of galaxies with over- and under-massive black holes are also distinguished by where the metals are stored inside their inner 0.1R$_{vir}$. In other words, galaxies hosting over-massive SMBHs have most of these metals in the gas phase, while galaxies with under-massive SMBHs have the metals in this region locked in stars. 


\begin{figure}[ht!]
\vspace{0mm}
\centerline{\resizebox{1\hsize}{!}{\includegraphics[angle=0]{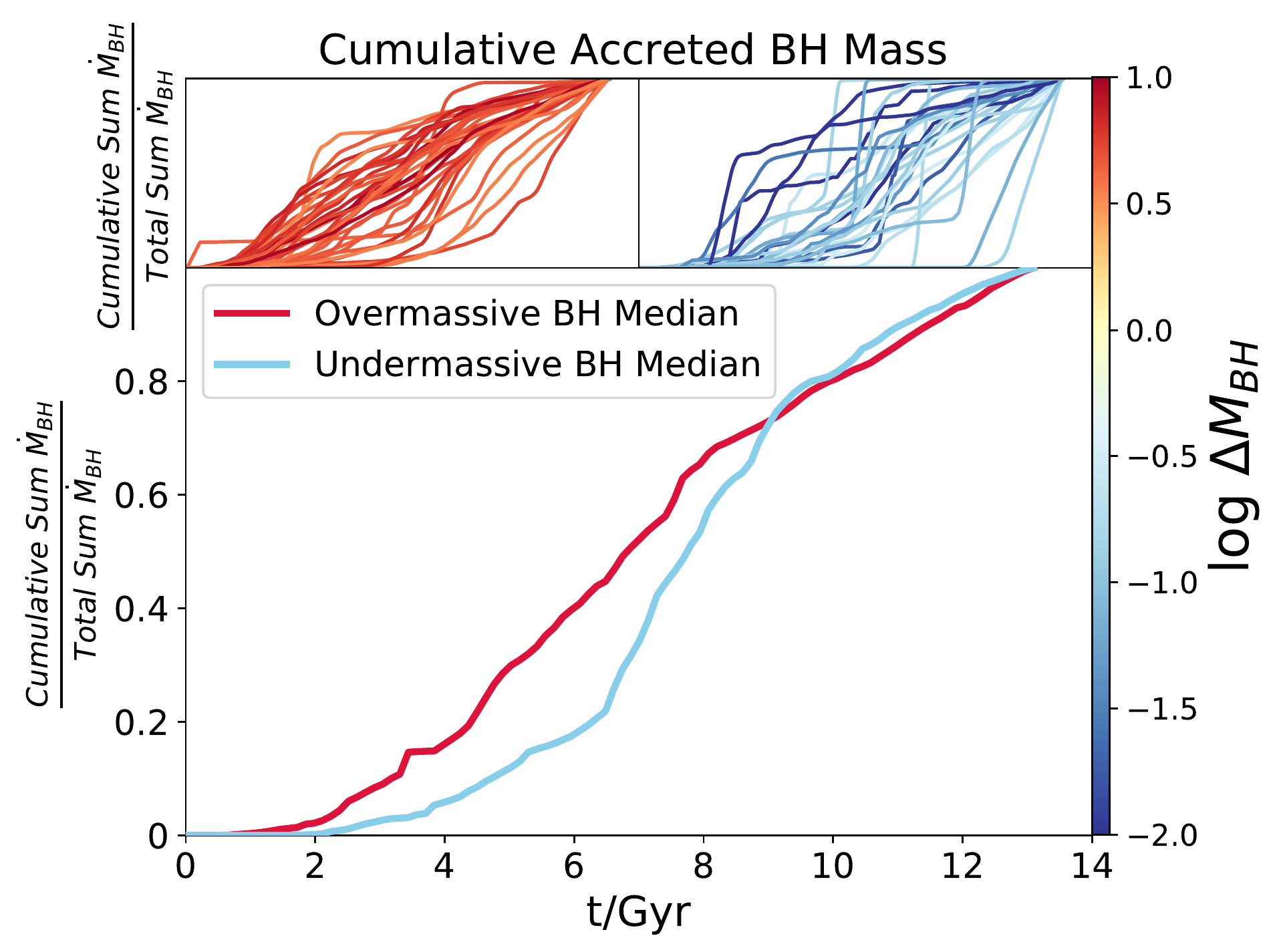}}}
\caption[]{The cumulative sum of the accreted black hole mass as a fraction of the total accreted BH mass as a function of time. \textit{Upper Panels:} Red and blue lines corresponds to this value for each individual galaxy with an over or under massive black hole, respectively. \textit{Lower Panel:} The red and blue line correspond to the median values for the galaxies with over-massive BHs and under-massive BHs, respectively. Here, we've only included black holes with $|\Delta M_{BH}| > 0.5$ to select for galaxies with the most significant mass deviation. We find that galaxies with over-massive BHs grow earlier than galaxies with under-massive black holes. \cite{Sharma2020} finds that the timing of galaxy growth is related to halo concentration.}
\label{figure:grad_accrBHmass}
\end{figure}

The right panel of Figure \ref{figure:fzcgm_fzdisk_fbaryons} shows the fraction of baryons in the CGM as a function of the baryonic fraction in the disk (points and colors as in left panel). The grey line indicates where the baryonic fraction of the disk and the baryonic fraction of the CGM equals one. Unlike the metal retention plot to the left, we see some scatter above and below this line. This difference is due to using the cosmic baryon fraction to calculate the expected $M_{total\ baryonic}$. 

Comparing the left and right panels in this figure, we see that the metals in the galaxy do not generally trace the baryonic component. When examining the metal retention plot, we see that most galaxies fall along or below the grey line and show a distinct trend with $\Delta M_{BH}$. Meanwhile, the baryonic fractions are quite different. Though the measurements are centered on the grey line, they otherwise have significant scatter above and below and the trend with $\Delta M_{BH}$ is not as pronounced. Nevertheless, galaxies with under-massive black holes (blue) tend to have more mass than expected and larger fractions of baryons in their disks (above the grey line) compared to galaxies with over-massive black holes which have baryon fractions below their expected masses and tend to have a higher fraction of baryons in their CGM. Thus, the galaxies can accrete more than their expected share of baryons or have reduced baryons due to lower accretion, rather than expulsion from its halo (which would result in lower metal retention rates beyond what we find). The differences we see between these plots tells us that the motion of the metals is not strictly following the motion of the gas and stars in the galaxy. 

Furthermore, we find that the galaxies with over-massive SMBHs are also those which are growing their SMBH through earlier accretion. Figure \ref{figure:grad_accrBHmass} (lower panel) shows the median values of our two subsamples for the cumulative sum of the accreted mass as a fraction of the total accreted mass for each central SMBH. The upper panels of Figure \ref{figure:grad_accrBHmass} show these values for each individual galaxy from which the median is calculated. We see that on average galaxies with over-massive black holes (red line) accrete material and grow their black holes at earlier times than their counterparts with under-massive black holes (blue line). This result is consistent with \cite{Sharma2019} and implies that there may be different avenues for the growth and formation of these SMBHs. Furthermore, \cite{Sharma2020} find that the different timescales of galaxy growth are related to their halo concentration.

\begin{figure*}[ht!]
\vspace{0mm}
\centerline{\resizebox{0.9\hsize}{!}{\includegraphics[angle=0]{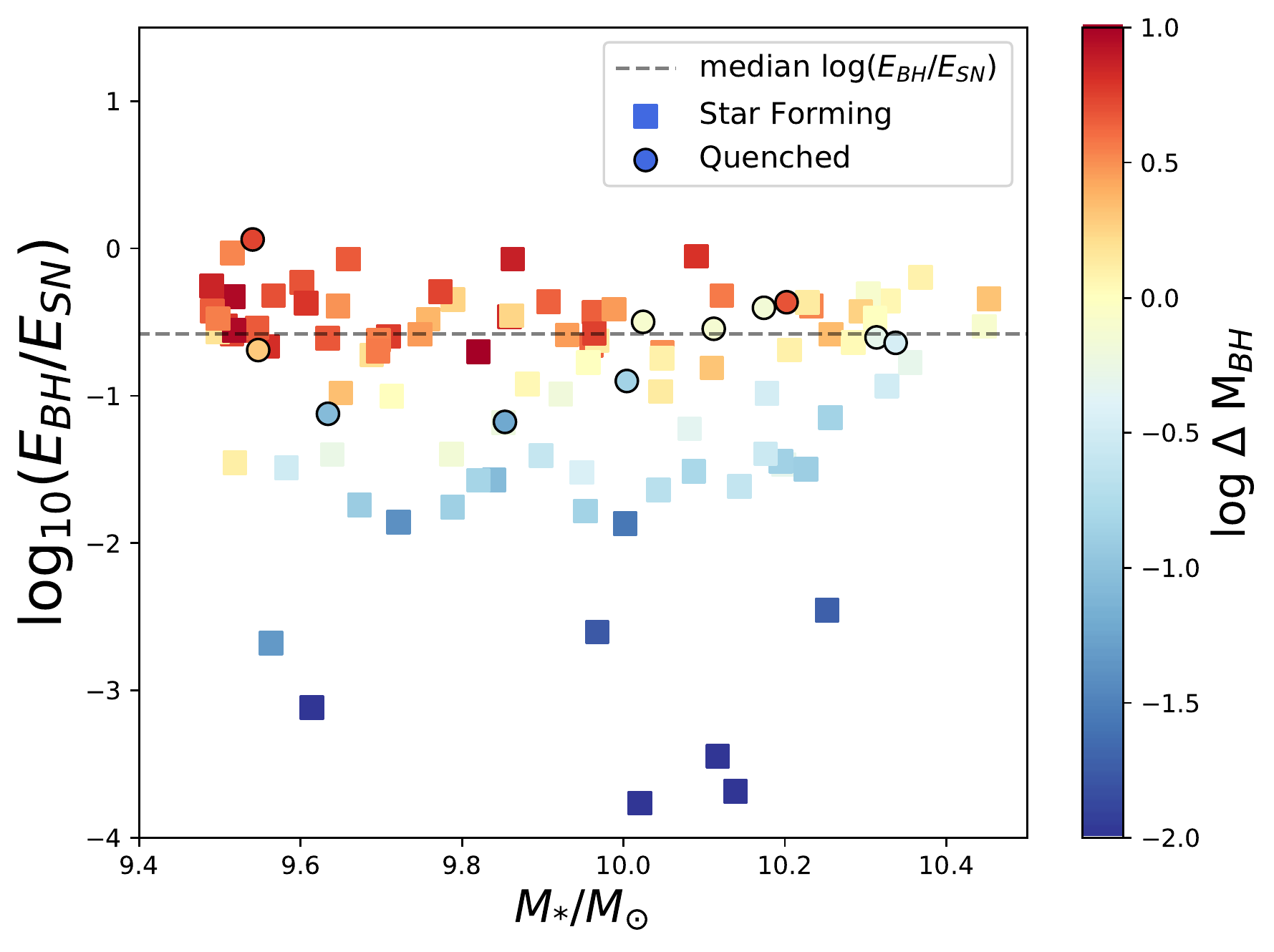}}}
\caption[]{The ratio of SMBH and stellar feedback energy as a function of stellar mass. Points are colored by $\Delta M_{BH}$. The dashed line indicates the median of the ratio of SMBH and stellar feedback energy, and galaxies that fall above this line have more AGN feedback than average. The galaxies with over-massive black holes (red) lie above this line even at the low mass end indicating that the impact of SMBH feedback is resulting in the differences in metal content that we see at the center of these galaxies.}
\label{figure:energetics}
\vspace{3mm}
\end{figure*}

\subsubsection{AGN Feedback Flattens Mass and Metallicity Gradients}

Within this lower mass half of our galaxy sample, we have shown that galaxies with differences in their SMBH mass residuals ($\Delta M_{BH}$) have different fractions of metals in the CGM. We measure the total energy output through SMBH feedback and stellar feedback within each galaxy and compare the ratio of these two quantities in Figure \ref{figure:energetics}. We find that the galaxies with over-massive SMBHs (red) experience more AGN feedback even at the low mass end where stellar feedback may have dominated. From Figure \ref{figure:energetics}, we argue that the feedback from SMBHs is responsible for impacting the local redistribution of mass in these galaxies over stellar feedback. Furthermore, we argue that SMBHs are more effective at redistributing mass in galaxies with comparatively smaller potential wells, which can explain the differences in metal retention that we see in Figure \ref{figure:fzdev}. 

\begin{figure*}[ht!]
\vspace{0mm}
\centerline{\resizebox{0.5\hsize}{!}{\includegraphics[angle=0]{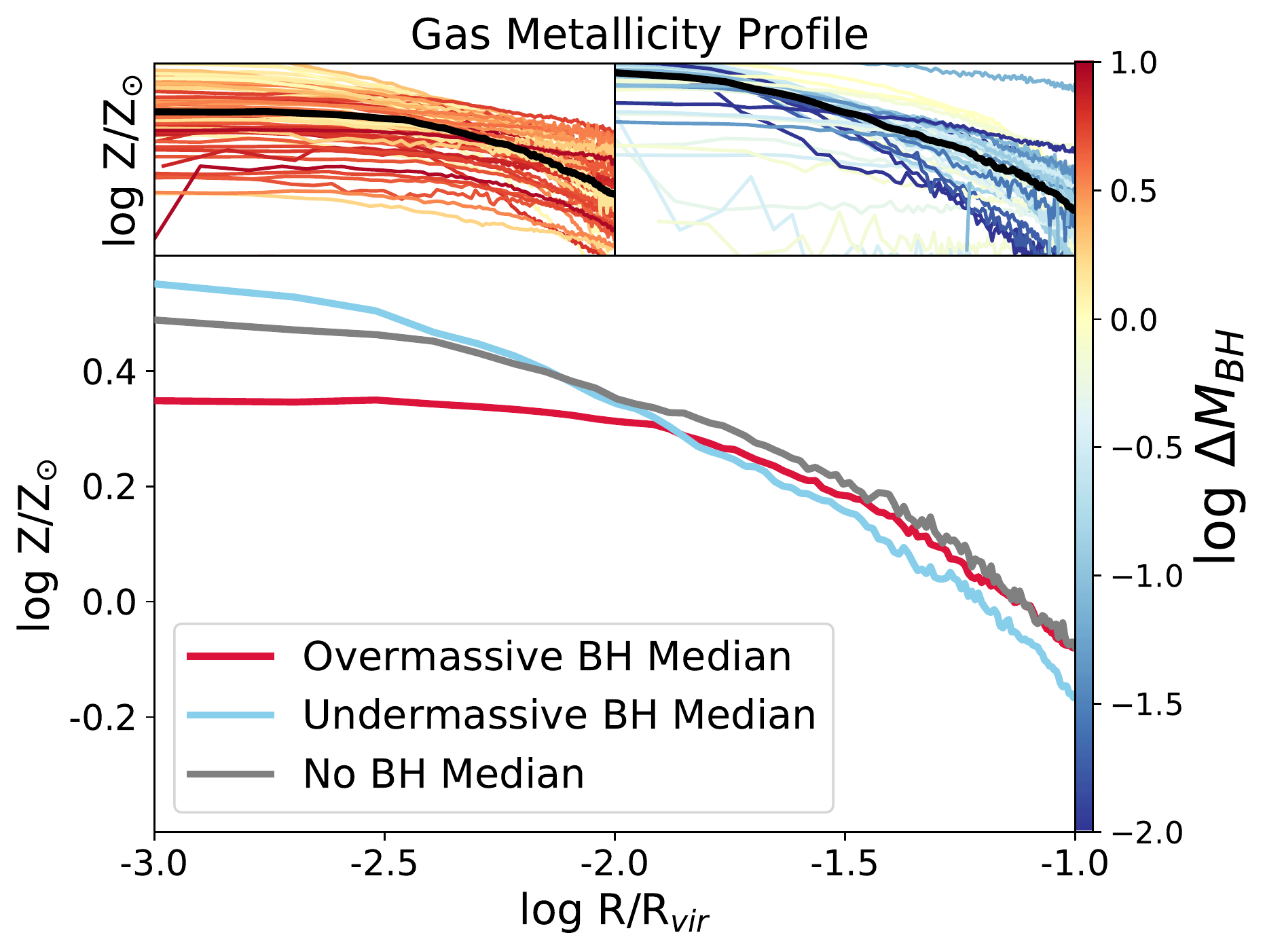}}
\resizebox{0.5\hsize}{!}{\includegraphics[angle=0]{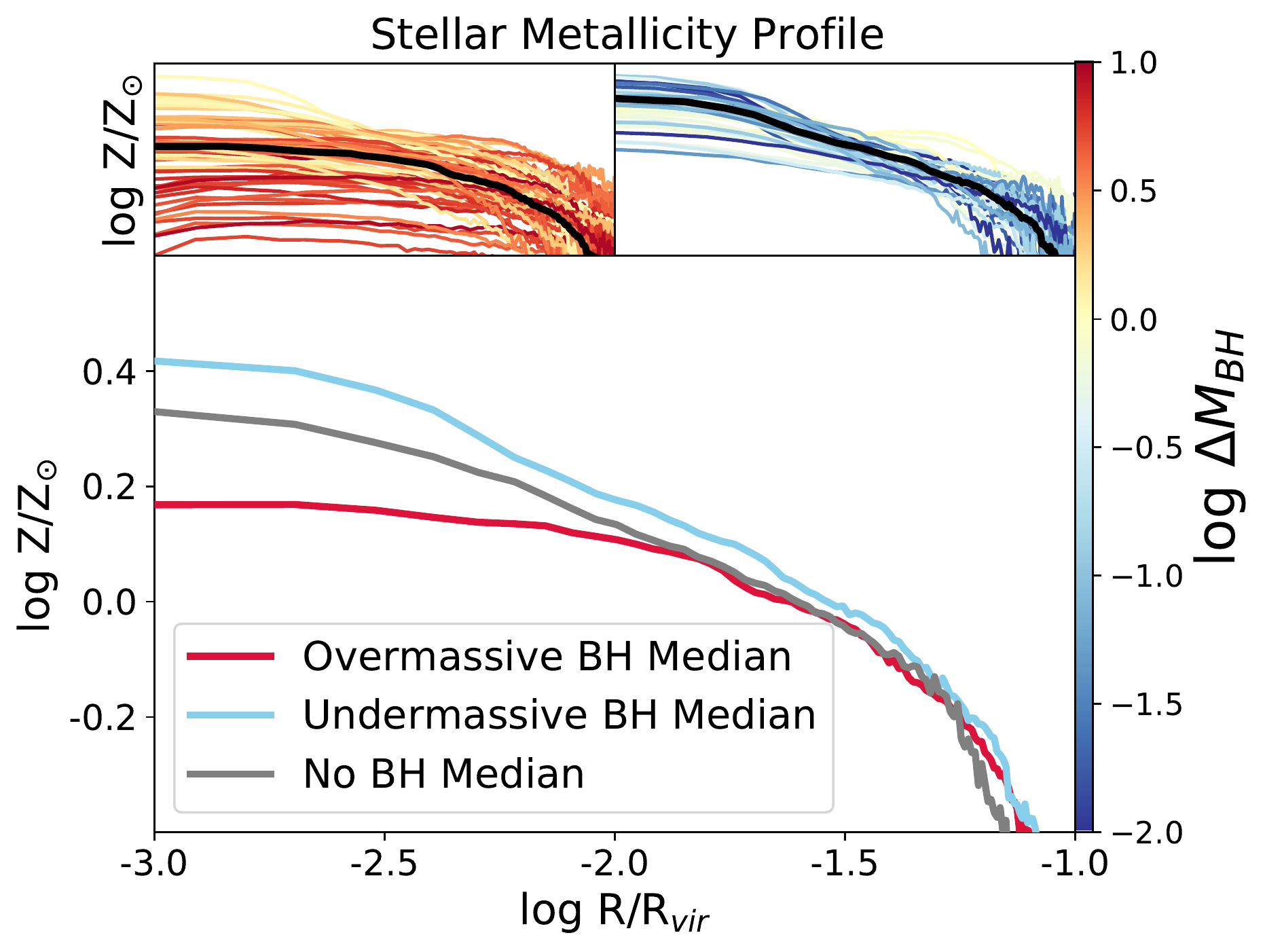}}}
\caption[]{Median gas-phase metallicity (\textit{Left}) and stellar metallicity (\textit{Right}) profiles for the galaxies that host over-massive (red) and under-massive (blue) SMBHs in our sample, including the median values for each subsample (bottom plots). In both pairs of profiles, galaxies with over-massive black holes show a flatter distribution of metals with no strong build up of metals in the center. By comparison, galaxies with under-massive black holes tend to have a build up of super-solar metal-rich gas at their centers and a steeper metallicity profile.}
\label{figure:gradZ_metals}
\end{figure*}

This process additionally impacts each galaxy's mass and metallicity gradients. We split the sample into two bins which included all the galaxies in the low mass sample with over-massive black holes ($\Delta$M$_{BH}$ \textgreater 0) and those with under-massive SMBHs ($\Delta$M$_{BH}$ \textless 0). Figure \ref{figure:gradZ_metals} shows the median gas phase (\textit{Left}) and stellar metallicity (\textit{Right}) profiles for all the galaxies with over-massive BHs (red) and those with under-massive BHs (blue). We also include the median metallicity gradients for the population of galaxies with no central SMBHs in grey. We see that, on average, galaxies with under-massive BHs or no BHs are more likely to have a concentration of metals built up within the gas and stars in their centers. By comparison, galaxies with over-massive BHs have fewer metals in their centers and have an overall flatter metallicity profile on average likely driven by the evacuation of metals by the SMBH. 

\begin{figure*}[ht!]
\vspace{0mm}
\centerline{\resizebox{0.5\hsize}{!}{\includegraphics[angle=0]{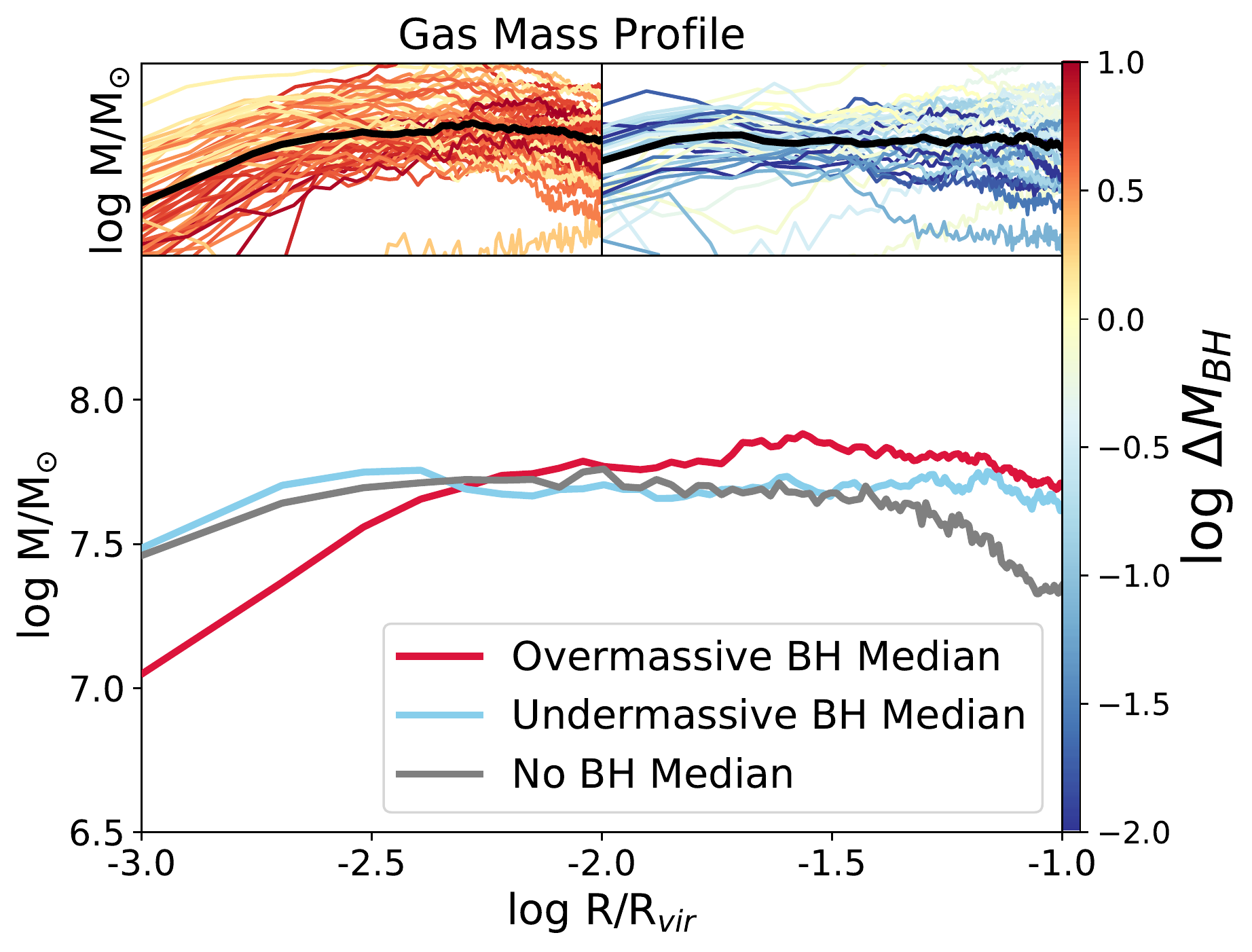}}
\resizebox{0.5\hsize}{!}{\includegraphics[angle=0]{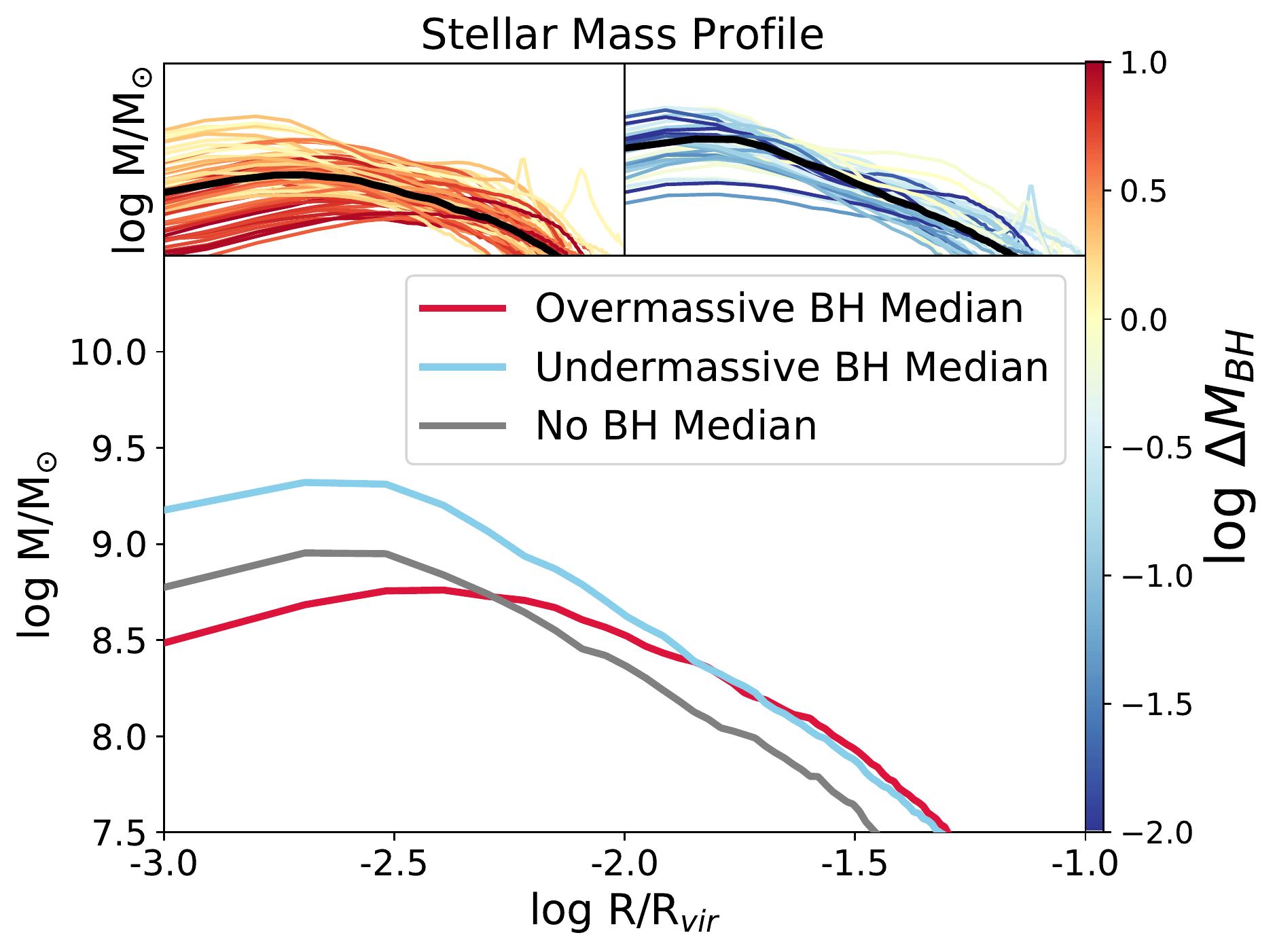}}}
\caption[]{Gas mass (\textit{Left}) and stellar mass (\textit{Right}) profiles for all the over-massive (red lines) and under-massive (blue lines) galaxies in our sample, including the median values for each subsample (bottom plot). Overall, the stellar mass profile for galaxies with both over- and under-massive SMBHs have a similar shape, but galaxies with under-massive black holes have about half a dex more mass in stars in their cores.}
\label{figure:gradZ_mass}
\end{figure*}

In Figure \ref{figure:gradZ_mass}, we additionally compare the gas mass and stellar mass radial profiles for this split sample. We see a similar stellar mass profile shape for both the populations of galaxies with over- and under-massive black holes; however, the galaxies with under-massive black holes have a larger build up of stellar mass in their centers by about half a dex. In the gas mass profile, we find that both the galaxies with under-massive BHs and without BHs have very similar profiles. In the stellar mass profiles, the median profile for the galaxies without BHs is lower than the galaxies with under-massive BHs. However, we argue that the shape of the stellar mass profile for the galaxies without BHs is closer to that of the galaxies with under-massive BHs. Additionally, the profile for galaxies without BHs is lower as they are likely to be less massive given that the occupation fraction decreases with decreasing stellar mass.

These profiles help explain what we see in the galaxies with under-massive BHs. These galaxies have SMBHs which grow to smaller masses than expected for their stellar velocity dispersion/potential (Figure \ref{figure:Msigma}). These under-massive BHs are less effective at regulating star formation across the evolution of the galaxy which results in two of the characteristics we see: the build up of stellar mass and both stellar and gas-phase metallicity in the center of the galaxies. 

Thus, due to less feedback from the SMBH, more metals end up locked in both the the gas and stars at the center of the galaxies with under-massive SMBHs and more stars form over all. This point further confirms one of the results of \cite{Sharma2019} which finds that the hosts of under-massive black holes followed nearly identical evolutionary tracks to galaxies without black holes. This result is also consistent with \cite{Sanchez2019} which shows that galaxies without BH physics end up with a significant build up of metals in their cores without BH feedback to eject the metals from the center and suppress star formation. 


\subsubsection{AGN Feedback Does Not Evacuate Gas from the CGM}

\begin{figure}[ht!]
\vspace{+2mm}
\centerline{\resizebox{1.1\hsize}{!}{\includegraphics[angle=0]{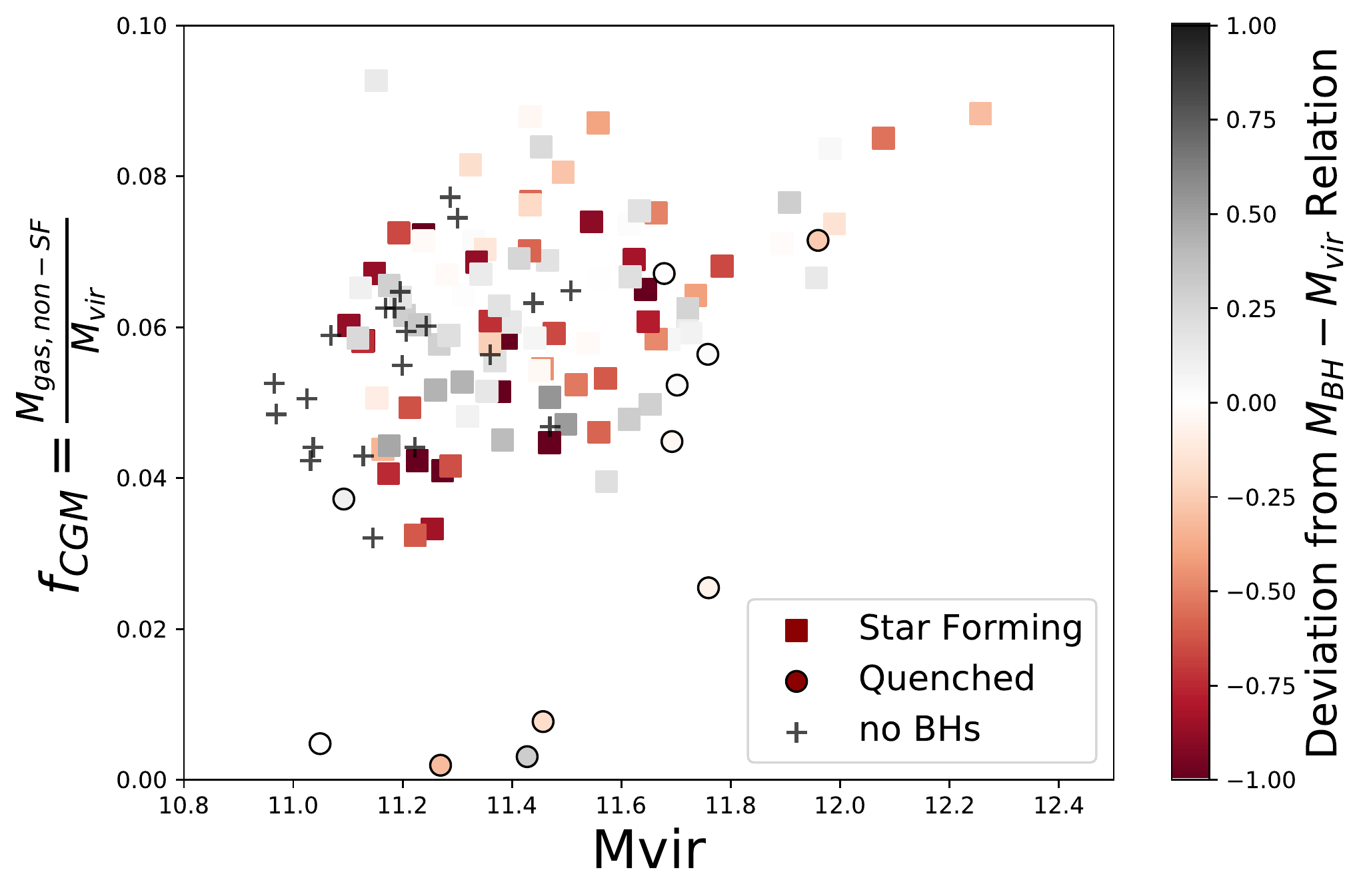}}}
\caption[]{ Measurements of the fraction of baryons in the CGM, as calculated in \cite{Davies2020}, as a function of the virial mass for all 140 of our galaxies. To compare more accurately with \cite{Davies2020}, points in this plot are colored by the deviation of each SMBHs mass from the rolling mean of the of M$_{BH}$-M$_{vir}$, $\Delta log_{10} M_{BH}-M_{vir}$, relation. Unlike Figure 2 from \cite{Davies2020}, we find no trend between f$_{CGM}$ and $\Delta log_{10} M_{BH}-M_{vir}$ for our galaxies indicating that at fixed halo mass, the deviation of black hole mass does little to impact the content of the CGM. We find a few quenched galaxies with nearly fully evacuated CGMs (lower left circles) which is consistent with rapid quenching (\textless 1 Gyr) found by \cite{Sharma2022} in intermediate dwarf scales ($10^{9.3}$ \textless $M_{*}$ \textless $10^{10}$).}
\vspace{0.5 mm}
\label{figure:fcgm_Mvir}
\end{figure}

We have shown that the local redistribution of mass by the SMBH in our galaxy sample impacts the metal retention in the disk and the metallicity profiles of our galaxies. To understand the impact of AGN feedback on the CGM of these galaxies, we compare our results to those of \cite{Davies2020}, which find that galaxies with over-massive BHs have a lower fraction of baryons in their CGM due to the evacuation of gas by black hole feedback both for IllustrisTNG galaxies and those from the EAGLE simulation. Interestingly, we find the opposite. Figure \ref{figure:fcgm_Mvir} shows the fraction of baryons in CGM gas, defined as in \cite{Davies2020}: 
\begin{equation}
f_{CGM} = \frac{M_{CGM}}{M_{vir}}
\end{equation}
where M$_{CGM}$ is the gas mass within M$_{vir}$ that is not star forming. Interestingly, we find no correlation with f$_{CGM}$ and the over- or under-massive state of the SMBHs from our galaxies.

This distinct difference between our findings and those of \cite{Davies2020} are likely due to the differences in the implementation of sub-grid BH physics. In the EAGLE simulation, AGN feedback \citep{2009MNRAS.398...53B} is implemented via stochastic, isotropic heating applied to gas particles ($\Delta$T$_{AGN}$ = 10$^{8.5}$ K) and the AGN feedback efficiency was chosen to reproduce the z = 0 scaling relation between galaxy stellar mass and their central SMBH masses. The energy injection rate is: 
$f_{AGN}\dot{m}_{acc}c^2$
where $\dot{m}_{acc}$ is the BH accretion rate and $f_{AGN}$ = 0.015 is a fixed value where $f_{AGN}^2$ determines the fraction of available energy coupled to the ISM. AGN feedback is the primary form of self-regulation in EAGLE once a hot CGM has formed, limiting the impact of stellar-driven winds out of the galaxy \citep{2017MNRAS.465...32B}.

In IllustrisTNG, AGN feedback is implemented in two modes: high accretion rates drive a feedback mode which injects energy thermally, heating nearby gas cells to the BH using an efficiency of $f_{AGN,thm}$ = 0.02, which is similar in scheme to our implementation in Romulus25. Meanwhile, feedback associated with low accretion rates inject energy kinetically with a random direction chosen for each injection event. The efficiency of the kinetic mode, $f_{AGN,kin}$, scales with local gas density up to 0.2, and kinetic AGN energy is injected with a velocity determined by the mass of gas associated with the inject region. The threshold between high and low accretion scales as a function of the BH mass and is written in terms of the Eddington ratio: 
\begin{equation}
\chi = min[0.1,\chi_0(m_{BH}/10^8 M_{\odot})^2]
\end{equation}
where $\chi_0$ = 0.002. Regardless of mass, a BH can inject feedback via the thermal mode at sufficiently high accretion rates \citep{Weinberger2017}; however, once a BH reaches the pivot mass of $10^8 M_{\odot}$, this mode becomes rare thereby setting this mass as the transition between thermal and kinetic feedback modes. 

While the simulations discussed in \cite{Davies2020} find that the SMBHs effectively eject gas from the CGM out into the IGM, our results are quite different. We find that in our galaxies the quantity $f_{CGM}$ does not appear to be influenced by $\Delta M_{BH}$. Thus, the majority of the SMBHs in our galaxies do not appear to evacuate gas and metals from the CGM out past the virial radius. The impact of the SMBH's feedback in relation to the depth of its host galaxy's gravitational potential (as sampled by stellar dispersion) doesn't appear to affect the baryon content of the CGM. Instead, we find that the mass and metal redistribution of the SMBH feedback is primarily limited to the local vicinity of the galaxy. We note that the AGN feedback in {\sc Romulus25} has been shown to be more moderate than in other cosmological simulations \citep{Tremmel2018a,Chadayammuri2021,Jung2022}. It may be that in the absence of high temperature metal cooling, our AGN can drive galaxy-scale outflows that can effectively regulate star formation but are not powerful enough to evacuate the CGM. We elaborate on this idea and its implications further in Section \ref{sec:metalcooling}.

Nevertheless, the stark differences between the measurements of the baryonic content of the CGM between the two simulations and from other groups will be an exciting test for future observations that measure gas and metal abundances in the CGM and include dynamical measurements of SMBH masses. See Section \ref{sec-discuss} for additional discussion.

Interestingly, there appear to be a small population of galaxies between log M$_{halo}$ = 10$^{11-11.5}$ that have significantly lower values of $f_{CGM}$ (Figure \ref{figure:fcgm_Mvir} bottom left). These galaxies also stand out in the right hand panel of Figure \ref{figure:fzcgm_fzdisk_fbaryons} (Blue circles at bottom center), and we see that these galaxies have $\Delta M_{BH}$ \textless \ 0. These galaxies have nearly all of their metal and baryonic content within the inner $0.1R_{vir}$ locked in stars, with little gas remaining in their center. \cite{Sharma2022} explored the quenching in these and lower mass dwarfs in {\sc Romulus25}. They find that galaxies with M$_* \sim 10^{10} M_{\odot}$, which are at the top of the dwarf mass regime, are more likely to experience quenching events that rapidly remove gas from the galaxy and partially evacuate the CGM within 1 Gyr. However, this characteristic effect appears to be confined to galaxies at or about $10^{10} M_{\odot}$, and not seen in galaxies above or below this mass. See \cite{Sharma2022} for additional details.




\begin{figure}[]
\centerline{\resizebox{1.0\hsize}{!}{\includegraphics[angle=0]{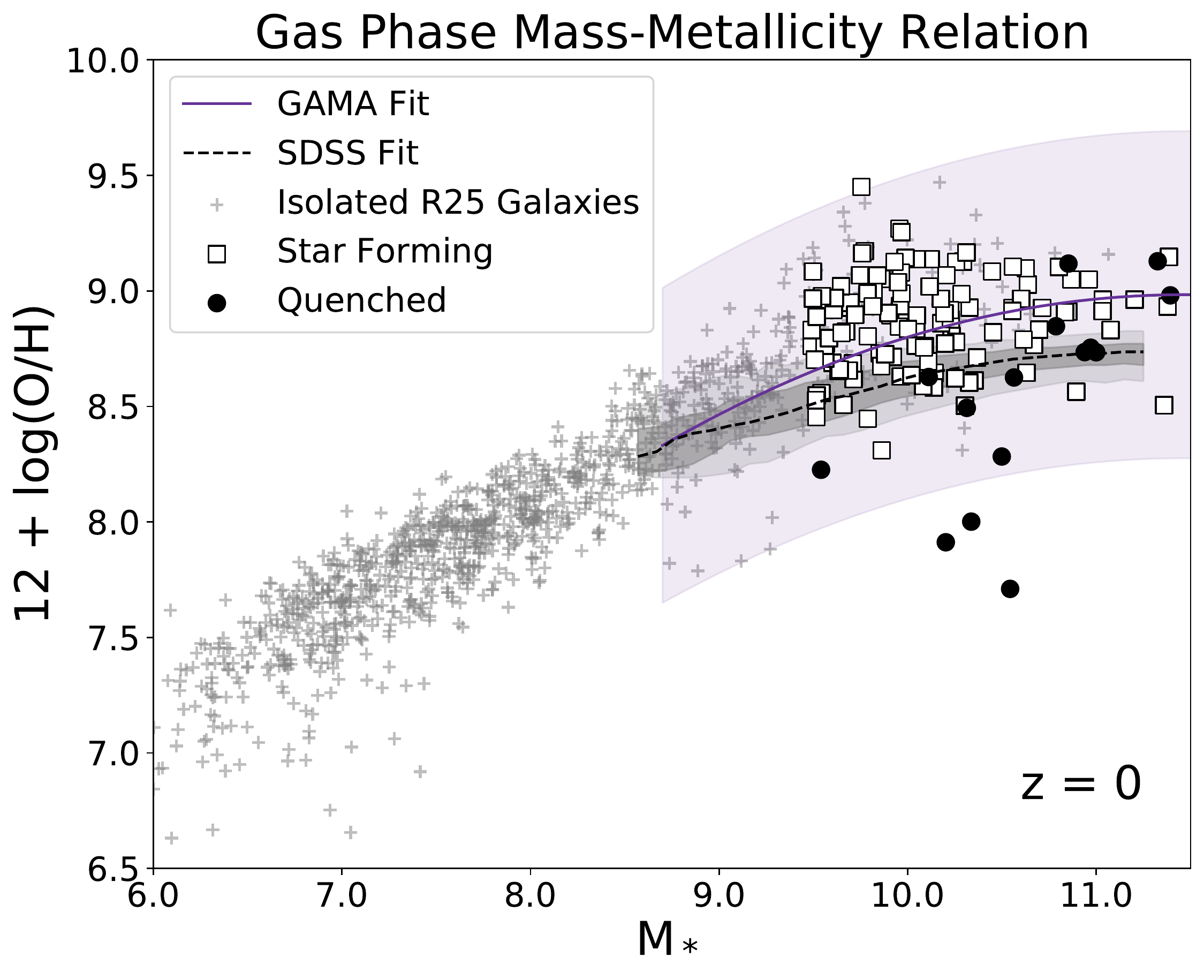}}}
\caption[]{The relation between the stellar mass and gas phase metallicity for our sample (black circles and white squares) and a wider selection of {\sc Romulus25} galaxies (grey crosses) at $z = 0$. The black dashed line indicates the SDSS fit \citep{Tremonti2004a,Pettini2004} and the purple solid line indicates the same relation from the Galaxy and Mass Assembly (GAMA) survey \citep{Foster2012} with their errors in shaded purple. Our sample of galaxies (3 $\times$ 10$^{9}$ \textless M$_{*}$ \textless 3 $\times$ 10$^{11}$ M$_{\odot}$)  fit well within the errors of the expected gas-phase metallicity of the galaxies from the GAMA survey, but over-predict the amounts expected from SDSS.}
\label{figure:MMR}
\end{figure}

\subsection{Predictions for Observational Comparisons}

To compare with both current and future observations, we create mock observations for metal retention measurements and predict \ion{C}{4} column densities for future surveys of quasar spectroscopy that include dynamical mass measurements. For this section of the analysis, we include only the star forming galaxies from our full galaxy sample due to the small number statistics of our quenched galaxies.

\subsubsection{Mock Observations of Metal Retention}
\label{subsec:fz_mockobs}

\begin{figure*}[ht!]
\centerline{\resizebox{1.1\hsize}{!}{\includegraphics[angle=0]{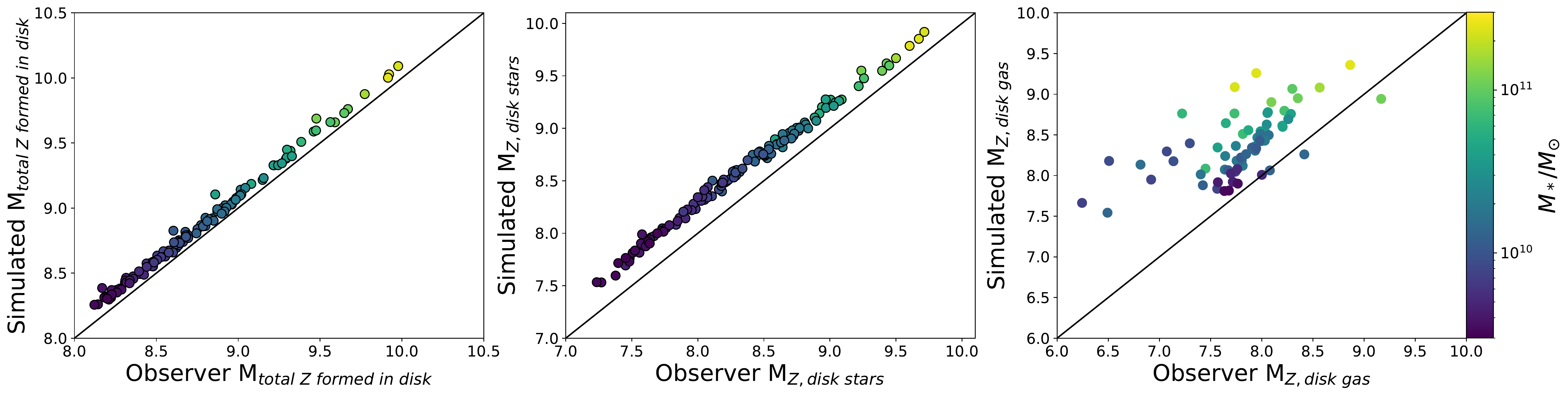}}}
\caption[]{Comparisons between the simulation measurement and tne mock observations for the $M_{total\ Z\ formed\ in\ disk}$ (\textit{Left}),  $M_{Z,\ disk\ stars}$ (\textit{Middle}), and $M_{Z,\ disk\ gas}$ (\textit{Right}). The mass of metals formed and metal mass in stars track each other well besides a small systematic offset due to the different definitions of the stellar disk in either case. The metal mass in disk gas however shows a more significant offset, as well as more scatter.}
\vspace{+5mm}
\label{figure:sim_obs_compare}
\end{figure*}

\begin{figure}[ht!]
\centerline{\resizebox{1.1\hsize}{!}{\includegraphics[angle=0]{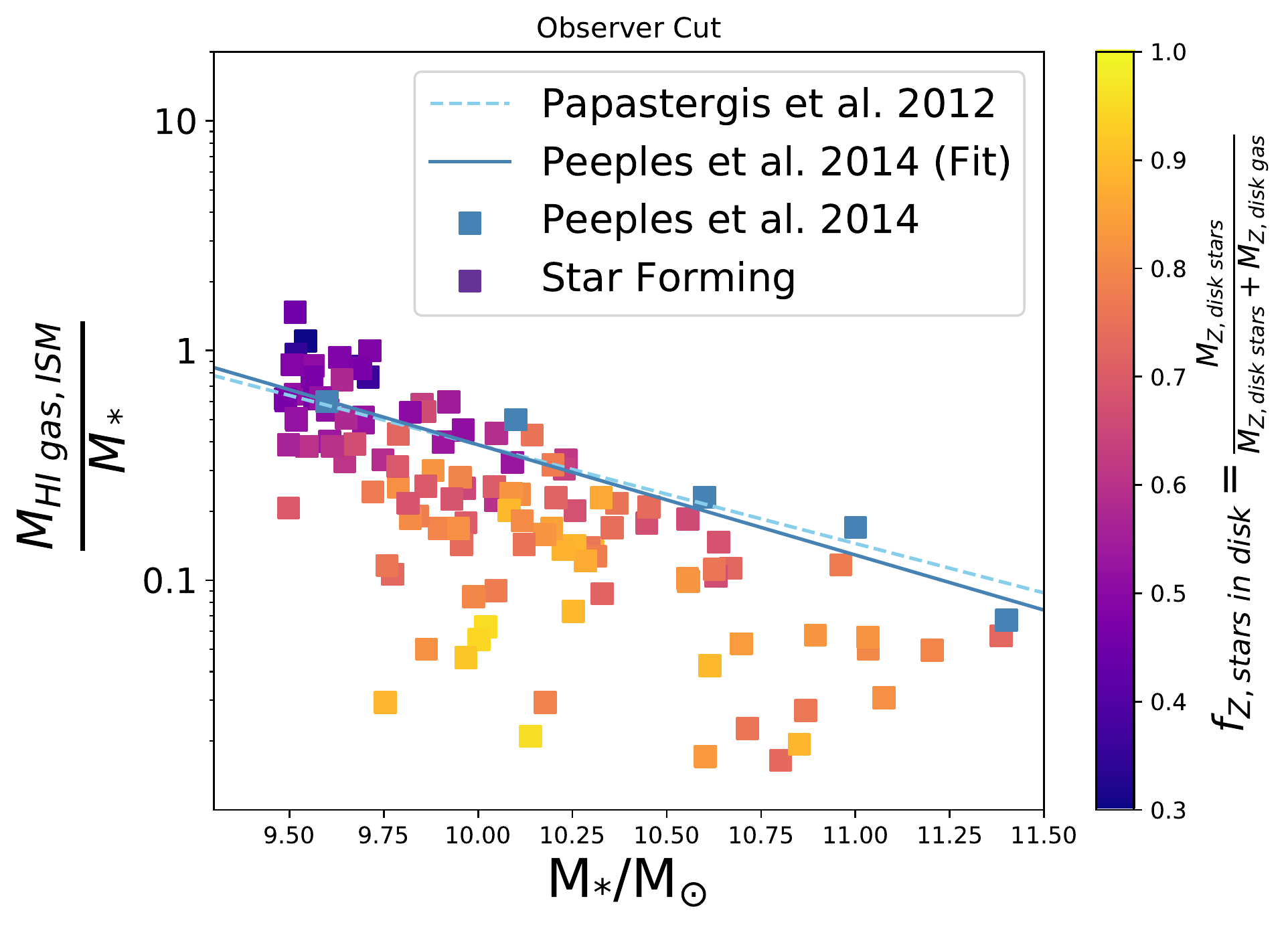}}}
\caption[]{The fraction of mass in HI gas to stellar mass within our star forming simulated galaxies as a function of their stellar masses. Observational constraints are included in dashed blue \citep{2012ApJ...759..138P} and solid blue lines and blue squares \citep{Peeples2014}. We find that our galaxies well match observation at the lower mass end of our SF sample; however, the HI gas mass fraction decreases to below expected values above M$_*$ \textgreater 10.25 $M_{\odot}$ and shows more scatter at this high mass end.}
\vspace{+5mm}
\label{figure:MHIgas_Mstar}
\end{figure}

Within the same mass range of the {\sc R25} galaxies analyzed in this work, past observations have calculated average metal retention fractions within the disks of selected nearby galaxies of between 20 \textemdash 40\% \citep{Peeples2014,Telford2019}. In comparison, the measurements for our simulated galaxies have an average of about 80\% as seen in Figure \ref{figure:fzdev} and Table \ref{table:medians}. To better compare to observations and determine whether these results could be observed, we create mock observations of the stars and gas in the simulated galaxy disks and process them in the method of an observer. This process results in a set of new values for the metal retention fraction for each galaxy. We follow the method of \cite{Telford2019} to calculate the necessary components of the metal retention equation (Equation 4) including the metal mass in stars in the disk, the metal mass in gas in the disk, and the metals produced by the stars in the disk.

Figure \ref{figure:MMR} shows the stellar mass and gas phase metallicity (Mass-Metallicity Relation, MZR) relation for galaxies selected from {\sc R25} to show the validity of this mock observation analysis. Grey crosses show galaxies across five orders of magnitude and the black circles and white squares indicate the same sample of galaxies we analyze throughout this paper. We find that the galaxies in {\sc R25} do follow the observed MZR of the GAMA Survey \citep{Foster2012} across the mass range of our galaxies, though our measurements are high compared to SDSS \citep{Kewley2008} depending on which calibration model is used to derive metallicities \citep{Tremonti2004a,Pettini2004}.

To move forward with our mock observations, we needed to construct ``mock observed disks'' from the simulated set of galaxies, including a stellar disk and gas disk. First, to select a stellar disk, we calculated the surface brightness of each galaxy and used the detection limit of SDSS (26 mags/arcsec$^2$) to define the radius and included all star particles within 1 kpc above and below the stellar plane. We use this definition of the stellar disk to calculate the SFH and the average stellar metal mass fraction in bins of 150 Myrs for each galaxy. These are the mock observables, since an observer could measure these quantities by modeling the observed light from the stellar populations. From these quantities, we calculate the metal mass locked into stars formed in each time bin as the product of the total mass of long-lived stars formed per bin \citep[assuming a returned fraction of R = 0.425;][]{Vincenzo2016} and the average metal mass fraction of the stars in that bin. The sum over all time bins then gives the total metal mass locked into disk stars, as an observer would measure.

We then use the same SFH and stellar metal mass fractions to calculate the metal production history for each galaxy, adopting the metal production model of \cite{Telford2019}. We do not attempt to match the metal production scheme in {\sc Romulus25}; rather, we opt to apply typical assumptions that an observer would make when calculating the metal mass produced by stars in an observed galaxy. We refer the reader to Section 3.1 of \cite{Telford2019} for details, but briefly summarize the key assumptions here. We account for metal production by Type II and Type Ia SNe, but exclude the contribution of AGB stars. For Type II SNe, we adopt metal yields from \cite{Nomoto2013} and account for the modest impact of stellar metallicity on predicted yields, and assume that only stars $\leq 40 M_\odot$ return metals to the ISM. For Type Ia SNe, we adopt the metal yields of \cite{Tsujimoto1995} and the delay-time distribution of \cite{Maoz2012}, and assume that the first Type Ia SN explodes 100 Myr after the onset of star formation. Using these ingredients, we construct a model of the metal production rate as a function of time following an instantaneous burst of star formation normalized to $1 M_\odot$ formed, then convolve this model with the “observed” SFH to calculate the metal production history. Finally, we integrate the metal production history over time to obtain the total metal mass produced by the stars in the “observed” disk for each galaxy.

Then, to select a gas disk, we used a surface density cut in HI of 10$^{17}$ cm$^{-2}$ and similarly included all the gas particles within 1 kpc above and below the disk plane. We note a caveat of our work: molecular hydrogen is not separately calculated due to the resolution of the simulation; therefore, the simulation HI includes all cold/cool ISM material \citep[see additional details in][]{Christensen2012}. From this gas disk selection, we calculated the total sum of HI mass within the disk and the average metallicity of cold, dense disk gas ($T < 10^4 K$, $\rho > 0.2\ m_p\ cm^{-3}$). The simulation does not resolve individual HII regions, but in real galaxies, the same gas in which star formation is ongoing is ionized by young, hot stars to produce the nebular emission from HII regions that observers use to measure the gas-phase metallicity. The total metal mass in the gas phase is then simply calculated as the product of the total “observed” gas mass and the average metal mass fraction in the cold disk gas for each galaxy.

The final outputs are measurements for mock observations of the following quantities:
\begin{itemize}
    \item metal mass locked in stars, $M_{Z,disk\ stars}$;
    \item metal mass in the gas phase, $M_{Z,disk\ gas}$;
    \item the amount of metals produced from the stars, $M_{total\ Z\ formed\ in\ disk}$;
\end{itemize}

and from those values we calculate a metal retention fraction, $f_{Z,disk}$, as in Eq. \ref{eq:fz}.

The first three measurements are compared in Figure \ref{figure:sim_obs_compare} for the measurements directly from the simulation and the mock observation. In the left-most plot, we compare the total metals formed in the disk, $M_{total\ Z\ formed\ in\ disk}$, for the measurement from the simulation (y-axis) and the mock observation (x-axis) and see that these values are closely aligned. There are systematically more metals formed when calculated directly from the simulation, but this difference arises from the slightly different definitions of the stellar disk: 0.1R$_{vir}$ vs the SB cut at 26 mags/arcsec$^2$. Similarly, we see a difference in the middle plot of Figure \ref{figure:sim_obs_compare} that shows the metal mass in the disk stars, M$_{Z,disk\ stars}$, with the measurement from the simulation showing more metals in disk stars than the mock observations. Again, this difference comes from the variation in the definition of the stellar disk.

The right-most plot of Figure \ref{figure:sim_obs_compare} shows the quantity with the largest scatter. The simulated and mock observations of $M_{Z,disk\ gas}$ show significant scatter as well as a systematic offset. We determine that this offset is likely a result of using only the mass of HI gas rather than the total gas mass in the observer calculation. However, Figure \ref{figure:MHIgas_Mstar} shows that the amount of HI in the ISM of our galaxies are near observed values at the low mass end, though are lower at higher stellar masses. These low values of HI can be explained by referring back to the right-most panel of Figure \ref{figure:sim_obs_compare} which shows that galaxies with higher stellar mass lie farther from the 1-to-1 line in grey. This trend is a result of higher fractions of HII gas residing in the hotter halos of more massive galaxies and thus depleting the amount of HI at these higher masses. It is likely that the inclusion of a full treatment of metal cooling could impact these results by increasing the total amount of cold gas traced by HI in our galaxy disks \citep{Christensen2014}. 

\begin{figure*}[ht!]
\centerline{\resizebox{1.0\hsize}{!}{\includegraphics[angle=0]{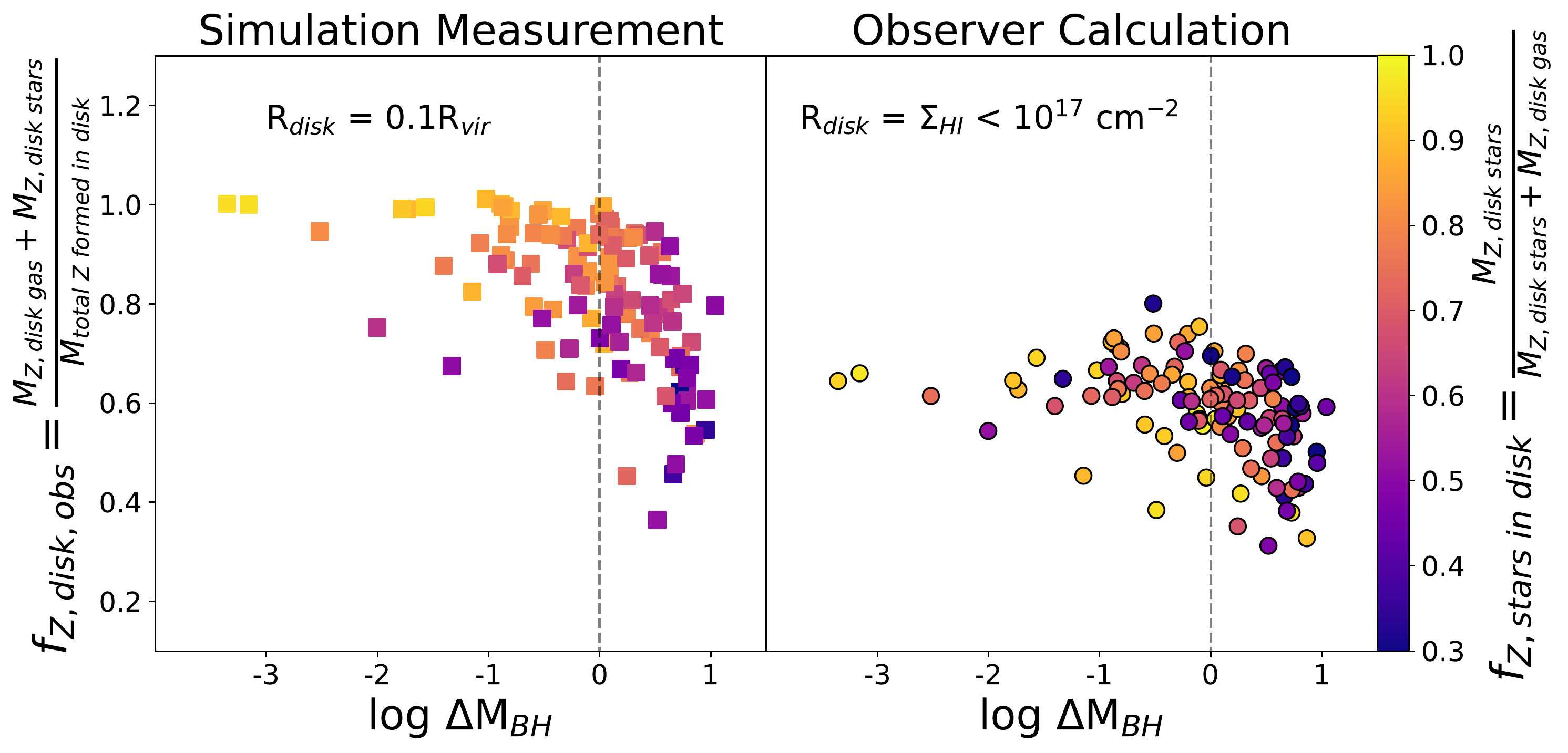}}}
\caption[]{The metal retention fraction from the simulation (\textit{Left}, as in Figure \ref{figure:fzdev}) and the mock observation (\textit{Right}). Despite a difference in the median metal retention fraction, our results hold for the mock observations: galaxies with over-massive black holes have lower metal retention fractions than galaxies with under-massive black holes. It is possible that this results could be tested by comparisons to metal retention observation that have associated black hole mass measurements.}
\vspace{+5mm}
\label{figure:fz_sim_obs}
\end{figure*}

Finally, we explore the impact of these differences by comparing the metal retention fraction, $f_{Z,disk}$, directly from simulation and for the mock observations in Figure \ref{figure:fz_sim_obs}. In this Figure (the left hand panel of which reproduces Figure \ref{figure:fzdev}), we compare the metal retention fractions as functions of log $\Delta$M$_{BH}$ with each point colored by the metal fraction of the disk/central region in stars for the simulated and mock observed quantities. 

Both panels of Figure \ref{figure:fz_sim_obs} have a qualitatively similar shape and show a flattened distribution below $\Delta$M$_{BH}$ $\lesssim$ 10 $M_{\odot}$ though with significant scatter. However, the relation peaks at $\sim$0.7 for the mock observation on the right instead of 1 as in the left hand panel. In both cases, most of the galaxies with the highest metal retention fractions and that host the most under-massive SMBH maintain more of their metals in stars (yellow points). Galaxies that maintain the most metals in the gas phase of their disks are primarily galaxies with over massive black holes (purple points on the right side of each panel). However, while the overall trend that under-massive black holes retain more of their metals in their disk holds for the mock observations, we note that the trend with metals locked in stars is present but not as clear.

Nevertheless, the relationship between $\Delta$M$_{BH}$ remains broadly consistent (i.e. over-massive black holes result in lower metal retention in the disk). Thus, even with biases inherent to the observational methods, there is potentially an observable separation between galaxies with high f$_Z,disk$ that host under-massive black holes compared to those with over-massive black holes with lower f$_Z,disk$ values. However, the trend between galaxies with metal rich disk gas and different $\Delta$M$_{BH}$ values is no longer as clear within the mock observations.

\subsubsection{Column Densities of \ion{C}{4}}

In addition to predicting how observable this effect may be through mock observations of metal retention, we additionally make predictions for how likely we might observe this phenomenon through quasar spectroscopy.

To do so, we calculate the column densities of \ion{C}{4} within the low mass SF galaxy sample. We choose to measure \ion{C}{4} because it acts as a viable proxy for $f_{CGM}$ as it traces metals within  T = 10$^4$\textemdash10$^5$ K and $\rho$ = 10$^{-5}$\textemdash10$^{-3}$ cm$^{-3}$ at $z=0$ \citep{Schaye2003,Oppenheimer2006,Dave2007,Ford2013,Rahmati2016,Oppenheimer2020}. Then we subdivide the SF galaxies into the two sets: galaxies with under-massive black holes and galaxies with over-massive black holes. Finally, we select matching pairs from each set with the closest available stellar masses to eliminate bias due to differences in stellar mass. This selection resulted in a subsample of 62 galaxies, 31 each with over- or under-massive SMBH, from which we could specifically isolate differences due to the SMBH mass excess. 

For each of these 64 galaxies we then calculated the \ion{C}{4} column densities, N \ion{C}{4}, of their inner 0.1$R_{vir}$. Column densities of \ion{C}{4} are calculated using the publicly available analysis software {\sc pynbody} \citep{pynbody}. Oxygen and iron enrichment from supernova and winds are traced throughout the integration of the simulation and carbon abundances are calculated from \cite{Asplund2009}. Ionization states are calculated during post-processing and assume optically thin conditions, collisional ionization equilibrium and an ultraviolet radiation field at $z=0$ \citep{Haardt2012}. Finally, we create models with variable temperature, density, and redshift using the CLOUDY software package \citep{2013RMxAA..49..137F} to determine the fraction of \ion{C}{4} in each gas particle.

\begin{figure*}[ht!]
\centerline{\resizebox{0.5\hsize}{!}{\includegraphics[angle=0]{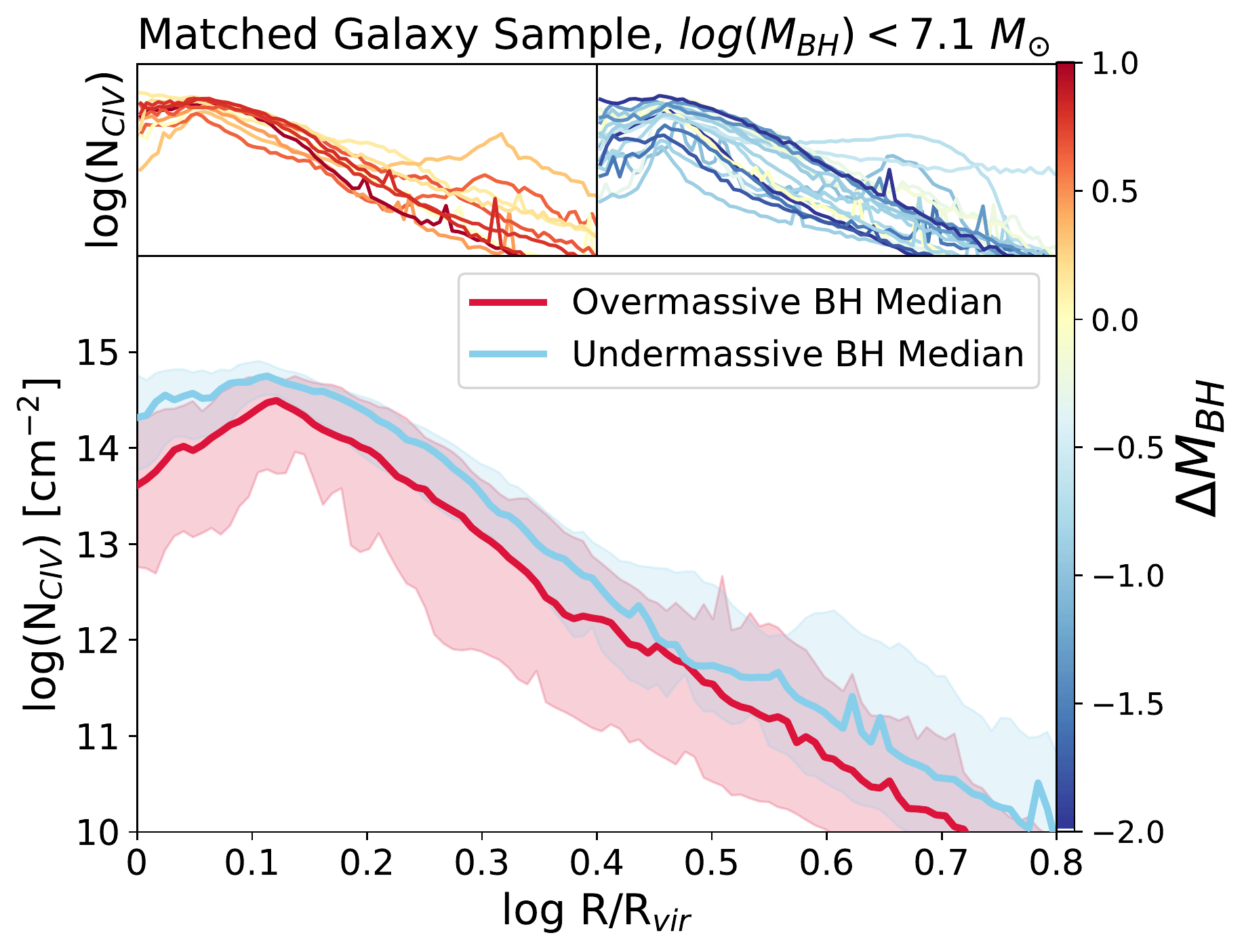}}
\resizebox{0.5\hsize}{!}{\includegraphics[angle=0]{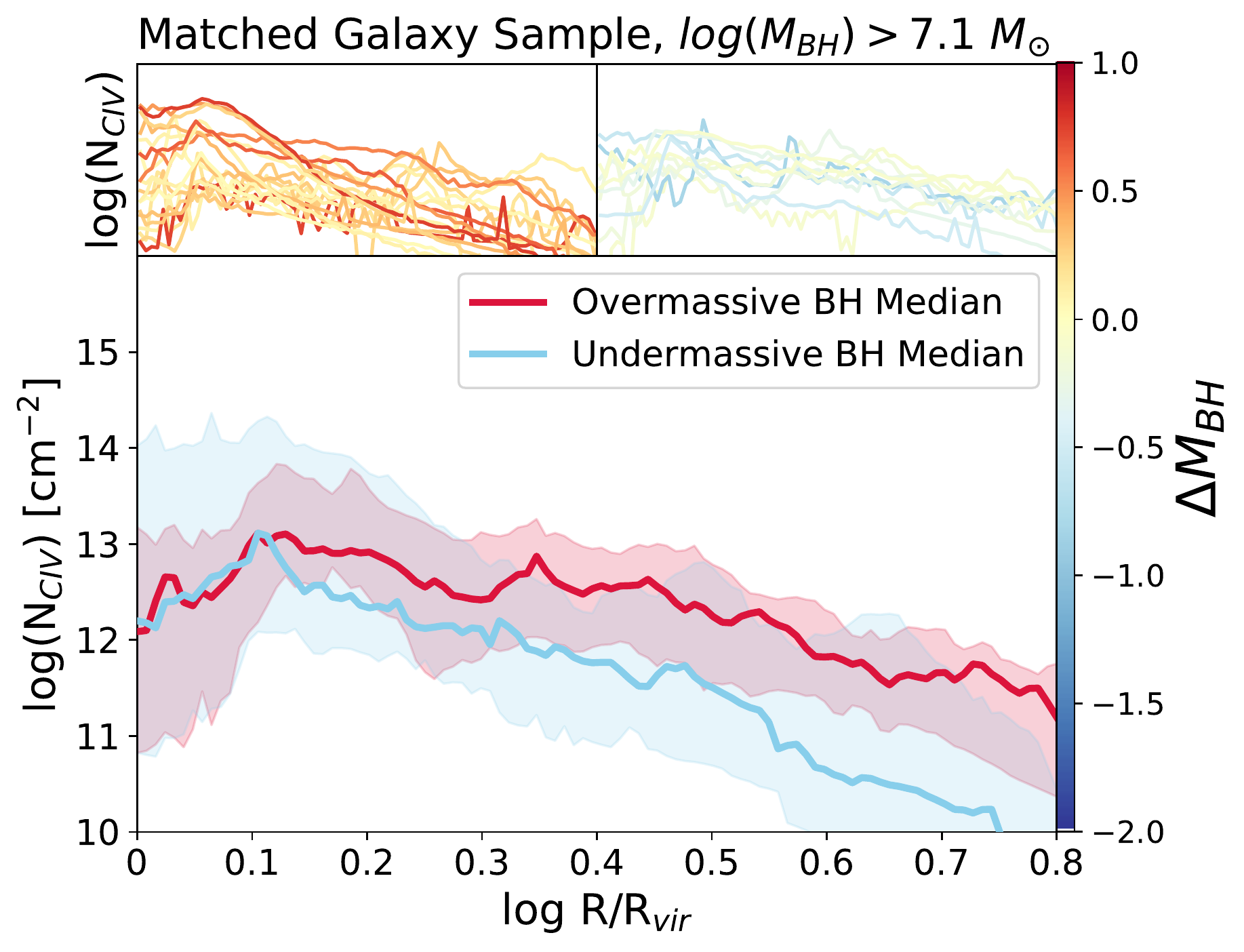}}}
\caption[]{Column densities of \ion{C}{4} as a function of radius for sub-sample of stellar-mass-matched galaxies with over-massive (red) and under-massive (blue) SMBHs. Shaded regions indicate the 16th to 84th percentile values. \textit{(Left):} Galaxies from our matched sample with black hole masses below 10$^{7.1}$ M$_{\odot}$. The upper panels show N(\ion{C}{4}) for each subset of galaxies split between over- and under-massive SMBHs and the lower panel shows the median for each subset. \textit{(Right):} The same set of figures as in the left panel but for galaxies with SMBH masses above 10$^{7.1}$ M$_{\odot}$. Our measurements predict that upcoming UV absorption missions that include host galaxy SMBH information, such as the COS-Holes survey, may observe a difference in the amount of \ion{C}{4} in their CGM. In galaxies with higher black hole masses, galaxies with over-massive black holes may contain significantly more \ion{C}{4} in the outer CGM, while a smaller difference in \ion{C}{4} content may be seen in galaxies with lower mass black holes.}
\vspace{+5mm}
\label{figure:grad_CIV}
\end{figure*}

From our 62 stellar-mass-matched galaxies, we split the sample into galaxies with SMBH masses above and below the median mass of $log_{10}(M_{BH}) = 7.1$, then we plot the column densities of \ion{C}{4} as a function of radius and compare the median values in Figure \ref{figure:grad_CIV}. In galaxies with lower mass BHs (left panel), there appears to be higher column densities of \ion{C}{4} in galaxies hosting under-massive BHs, with a consistent difference of about 0.3 dex throughout the content of the CGM. In galaxies with higher mass BHs (right panel), we find that there is more abundant \ion{C}{4} in the CGM of galaxies with over-massive SMBHs (red lines) than those with under-massive SMBHs (blue lines). Each shaded region indicates the 16th to 84th percentile value for each subsample. The difference between the median values of N\ion{C}{4} appears to increase between 0.1 to 0.8 R$_{vir}$. However, when we compare the entire sample of 62 matched galaxies in this way, the differences between column densities of \ion{C}{4} disappear. 

Future HST/COS surveys will be able to determine whether these predictions are borne out by observations. COS-Holes \citep{COS-HOLES}, one such future survey, will connect the UV absorption measurements made with COS to dynamically-resolved SMBH measurements for galaxies in the line-of-sight of the background quasar. We discuss the implications of this result and compare our predictions to those from other simulations in Section \ref{subsec:thefutureofNciv}.

\section{Discussion and Conclusions} 
\label{sec-discuss}

\subsection{Metal Cooling May Erase Evidence of AGN Outflows}

\label{sec:metalcooling}
The lack of metal cooling in {\sc Romulus25} is a major caveat of our results. Lack of metal-line cooling will result in an  underestimate of cooling rates of hot (10$^{5-6}$ K) gas by a factor of a few \citep{Shen2010}. If metal cooling were included, it is possible that metal enriched gas put into the CGM from AGN outflows may cool more rapidly and make their way back onto the central galaxy. The lack of metal cooling in Romulus may artificially enhance the signature we see (less metals retained in the galaxy and more put into the CGM) while in reality this signature may be more transient.

However, we argue that our key result remains valid. The relationship between a SMBH's mass and its host's galactic potential (from sigma by proxy) does determine how effective SMBH feedback is at driving outflows of metal enriched gas from its host galaxy. For example, the galaxies with over-massive BHs will have their metals expelled from the galaxy and into the CGM; however, when metal cooling is included, the high fractions of metal-enriched CGM gas (as seen in Figure \ref{figure:fzcgm_fzdisk_fbaryons}) may not be observed at z $\sim$ 0 due to faster cooling times that allow that gas to cool back onto the galaxy more rapidly.

While the length of time that metal enriched gas remains in the CGM may be overestimated by the cooling treatment in Romulus, our main result that scatter within the $M-\sigma$ relation is connected to the thermal and kinematic histories of metal enriched gas still holds, with overly massive black holes more likely to drive larger outflows of metal enriched gas further into the CGM.


\subsection{BHs Primarily Drive Metals into the CGM not the IGM}
\cite{Sharma2019} looked at a sample of 205 isolated dwarf galaxies with central SMBHs from the {\sc Romulus25} volume. They determined that galaxies with over-massive black holes formed earlier and exist in galaxies with lower stellar masses than expected for their halo. Despite exploring a higher mass regime of galaxies, we see similar indications of lower stellar masses at the centers of our galaxies that host over-massive black holes indicating that the BH may play a role in suppressing the integrated overall star formation of the galaxy. These results are also consistent with \cite{Davies2020} who found that in a sample of galaxies from EAGLE and IllustrisTNG those with over-massive black holes formed within dark matter halos with tightly bound centers and were characterized by systematically lower star formation rates. 

However, \cite{Davies2020} find that the galaxies in their sample with over-massive BHs negatively correlate with the fraction of total gas in their CGM. In other words, galaxies that contain more massive BHs at fixed galaxy mass evacuate more of their CGM. In contrast, the total baryon fractions in the CGM of our galaxies do not correlate with the deviation of mass of our SMBHs. Figure \ref{figure:fcgm_Mvir} shows no trend with $\Delta M_{BH}$ so the over-massive black holes in our simulated galaxies are \textit{not} evacuating the CGM as is the case for EAGLE and IllustrisTNG. Instead, the trends that we see imply that the over-massive BHs in our galaxies may be effective at redistributing both gas and especially metals within the disk/central region  ($\sim 0.1R_{vir}$) but they are not adept at evacuating material out of the CGM and into the intergalactic medium, except in a few, specific cases for galaxies with $M_* = 10^{10} M_{\odot}$ \citep[Figure \ref{figure:fcgm_Mvir}][]{Sharma2022}. The difference between our results and \cite{Davies2020} is likely a result of the different implementations of BH feedback we discussed in Section \ref{sec-results}. 

From Figure \ref{figure:energetics}, we confirm the SMBH's role in driving metals out of the center of the galaxy. This result is also consistent with the work of \cite{Sanchez2019}, which explored the metal content of the CGM in a set of 4 galaxies run with the same code and nearly the same resolution ($M_{gas}$ = 2.1 $\times$ $10^5$ $M_{\odot}$ , $M_{DM}$ = 1.4 $\times$ $10^5$ $M_{\odot}$)  to {\sc R25}. In that paper, they compare 4 galaxy simulations with and without black hole physics and find that the galaxies without BH physics end up with a concentration of metals in their centers as we see here in the galaxies with under-massive BHs. Their Figure 10, which compares the metallicity profiles for ISM gas for the galaxies and shows a build up of metals in the center for the galaxies without black holes is quite similar to our Figure \ref{figure:gradZ_metals}. In our case, the galaxies with under-massive black holes show the same build up of metals at the center as the galaxies with no black holes which is consistent with \cite{Sharma2019}. The consistencies between these simulations further confirms that the SMBH, which was responsible for ejecting metals from the galaxies in \cite{Sanchez2019}, is also key in ejecting metals from the galaxies with over-massive BHs in our sample.


\subsection{Predictions for Future CGM Surveys}
\label{subsec:thefutureofNciv}

From Figure \ref{figure:grad_CIV}, we predict that surveys like COS-Holes \citep{COS-HOLES} may see a distinction between the amount of \ion{C}{4} and other ions in the CGM of galaxies that host over- and under-massive central SMBHs. An observable test for these predictions could come in the form of future HST/COS observations, specifically those like COS-Holes, that will pair UV absorption measurements to dynamical BH mass measurements. With such observations, we will additionally be able to determine whether or not SMBH are evacuating gas in the CGM of their hosts. Furthermore, these kinds of metal line measurements, paired with dynamical BH mass estimates, would allow us to determine whether SMBHs that are over- or under-massive play different roles in setting the metal content of the CGM.

One compelling case is M31. \cite{Telford2019} measured the disk of the Andromeda Galaxy had lost up to 62\% of the metals formed by its stellar population. Therefore, the metal retention of the disk, $f_{z,disk}$, is 38\%, which is within the range of the metal retention rates we find in our sample using the observer method (\textit{Right}, Figure \ref{figure:fz_sim_obs}). The galaxies with the lowest metal retention rates nearly all host over-massive black holes, which is the case for M31. M31 has a velocity dispersion in the bulge of 151-153 km/s \citep{Whitmore1980,Zou2011} and a central SMBH mass of 1.4 $\times$ 10$^8$ $M_{\odot}$ \citep{Bender2005} which is 1.5$\times$ larger than expected based on equation \ref{eq:KandH} \citep{Kormendy2013}. 

While this is only one case, it demonstrates a clear example of galaxy with metals that have been ejected from the disk in the presence of a SMBH that is over-massive compared to its stellar dispersion. Our study shows that there is plenty of exciting work to be done connecting the flow of metals in a galaxy to the properties and effects of SMBHs. 







\section{Acknowledgements}

This work was supported by the FINESST19-23 grant 80NSSC19K1409 and NSF MPS-Ascend award ID 2212959. NNS and JKW acknowledge additional support for this work from NSF-AST 1812521, NSF-CAREER 2044303, the Research Corporation for Science Advancement, grant ID number 26842. Some of the predictions made in this study will inform the observational survey being carried out under HST-GO-16650, ``Connecting Galaxy Black Hole Mass with the State of the Circumgalactic Medium.” A collaborative visit was funded by the European Union’s Horizon 2020 research and innovation programme under grant agreement No. 818085 GMGalaxies. TRQ acknowledges support from Blue Waters and XSEDE, and the simulation in this study was run on NAS. Analysis was completed on NAS (under NASA award SMD-21-75544133). NNS gratefully acknowledges helpful conversations with the following individuals: Andrew Pontzen, Jonathon Davies, Martin Rey, Hannah Bish, Ben Oppenheimer, Ferah Munshi, Jillian Bellovary, Anna Wright, and Julianne Dalcanton.

\bibliography{./Sanchez2021_CGMZ.bib}



\end{document}